\documentclass[letterpaper, paper,10pt]{AAS}		

\usepackage{bm}
\usepackage{amsmath}
\usepackage[colorlinks=true, pdfstartview=FitV, linkcolor=black, citecolor= black, urlcolor= black]{hyperref}
\usepackage{overcite}
\usepackage{footnpag}			      	

\usepackage{amsfonts}
\newcommand{\norm}[1]{\left\|#1\right\|}
\usepackage{subcaption}
\usepackage{graphicx}
\usepackage{commath}
\usepackage{hyperref}
\renewcommand{\vec}{\boldsymbol}
\usepackage{color, colortbl}
\definecolor{Gray}{gray}{0.9}
\definecolor{LightCyan}{rgb}{0.95,1,1}

\newcommand{\Eqref}[1]{Eq.~\eqref{#1}}

\usepackage{floatrow}
\PaperNumber{23-242}

\begin{document}

\title{ORBITAL CONTROL STRATEGY FOR A CUBESAT SATELLITE EQUIPPED WITH A SOLAR SAIL FOR EARTH-MARS COMMUNICATIONS DURING SOLAR CONJUNCTIONS}

\author{Leonor Cui Domingo Centeno \thanks{MSc Student, Faculty of Mathematics, Complutense University of Madrid, Plaza de las Ciencias 3, 28040 Madrid, Spain.
\\ Research Intern, Departament of Geodesy, National Geographic Institute, Calle del Gral. Ibáñez de Ibero 3, 28003 Madrid, Spain. \texttt{leonorcd@ucm.es} },  
\ and Ariadna Farrés \thanks{Dr. Research Scientist, University of Maryland Baltimore County, 1000 Hilltop Cir, MD 21250, Baltimore, United States. \texttt{ afarres@umbc.edu}}
}

\maketitle{}

\begin{abstract}
This paper presents a mission concept that enables Earth-Mars communications resistant to periods of solar conjunction by using CubeSat satellites equipped with a solar sail. The dynamics of the satellite is modeled separately in the respective Earth-Sun and Mars-Sun Restricted Three Body Problem (RTBP), modified to include the solar radiation pressure effect exerted on the sail. Due to the non-linearities presented on this model, we numerically determine the location of the non-eclipsed equilibrium points parameterized by the sail orientation through a continuation method. These are the points where two CubeSat nanosatellites equipped with a solar sail could be placed. The instability of these equilibrium points makes it necessary to implement a control strategy to keep the satellite's trajectory close to equilibrium by constantly changing the orientation of the sail. To prove the robustness of the strategy, some numerical simulations have been performed for a given period of mission.
\end{abstract}
\section{Introduction}
In recent years, advances in space technology have allowed a greater and deeper study of Mars, encouraging the development of new missions\cite{Johnson_MSR, Tang_2020, percy2004study}, nowadays becoming a constant in the space sector. In particular, communications between Earth and Mars when both planets are eclipsed by the Sun have been a challenge in the past decades. In this paper, we propose a novel mission concept that can successfully address these planetary conjunctions (which occur every two years) \footnote{\url{https://www.nasa.gov/feature/jpl/whats-mars-solar-conjunction-and-why-does-it-matter}}. Showing that it can be accomplished with two CubeSat nanosatellites equipped with a solar sail. 
\\\\
A solar sail is a large orientable surface made of a lightweight and highly reflective material that allows to propel satellites by taking advantage of the momentum transfer between the incident photons from the Sun and the sail reflecting material. The solar sail is characterized by the orientation angles $\alpha$ and $\delta$ and the sail lightness parameter $\beta$. 
This way, we propose to use solar sails, which eliminate the need for high thrust propulsion systems, only requiring the inexhaustible source of energy from the Sun and combine them with CubeSat nanosatellites, which allow us to replace traditional larger-sized spacecrafts. The reduced size and weight of these satellites make the effect of the solar radiation pressure more relevant. Therefore, this approach aims to reduce significantly the mission costs and enabling long-term missions.\cite{Helvajian, villela2019towards, mcinnessolar}
\\\\
To the date, there have been several successful solar sail missions such as IKAROS\footnote{\url{https://global.jaxa.jp/countdown/f17/overview/ikaros_e.html}} by JAXA, NanoSail-D2\footnote{\url{https://www.nasa.gov/mission_pages/smallsats/nanosaild.html}} by NASA, and LightSail\footnote{\url{https://www.planetary.org/sci-tech/lightsail}} by The Planetary Society. In particular, the combined use of a 3U CubeSat satellite with a solar sail has been successfully tested by NanoSail-D2 and LightSail 2, where a solar sail was deployed in a low Earth orbit. 
\\\\
In this paper, given the small gravitational attraction between both planets, we study the Earth-Sun and Mars-Sun systems separately, both including their corresponding CubeSat satellite propelled by a solar sail. The dynamics of each problem can be modeled using the corresponding Restricted Three Body Problem (RTBP), augmented with the acceleration due to the solar radiation pressure effect exerted on the sail. 
\\\\
Due to the inherent non-linearities present in the satellite’s equations of motion, we start our study by determining the equilibrium points. However, as no explicit formula can be obtained, a continuation method is employed to find a surface of equilibrium point solutions parameterized by the sail parameters $\beta$, $\alpha$ and $\delta$. 
The continuation method used consists of iteratively obtaining the equilibrium points through a Newton-type method, performed along the discretization of a parameter's domain, as will be described in this paper. Then, for a fixed value of $\beta$, 5 continuous families of equilibrium points $FL_i$ with $i \in \lbrace 1 \dots 5 \rbrace$, are obtained in the $xy$ plane while only 3 families of equilibrium points, $FL_{1,2,3}$, are obtained in the $xz$ plane for different values of the sail parameters. 
\\\\
For the desired satellite communications between Earth and Mars during the above mentioned planetary conjunctions, we focus on the family of equilibrium points $FL_1$ in both systems given that it is closer to the Sun. Moreover, the equilibrium points of $FL_1$ that rise above the ecliptic plane will be of our interest, since they will not be eclipsed during the conjunction events and allow communication between Earth and Mars. The stability and classification of these points based on its linear dynamics are also analyzed in order to place the satellites in a feasible non-eclipsed equilibrium point of $FL_1$ for both systems. It is found that for both planets the admissible non-eclipsed equilibrium points are unstable, with a classification of type saddle $\times$ center $\times$ center. Hence, the main instability is due to the existence of an unstable manifold associated to a saddle behavior and makes it necessary to implement an orbital control strategy for remaining close to these equilibrium points.
\\\\
In this paper we present a control strategy that prevents the satellite from drifting away from the nominal position, thus avoiding communication cuts. This strategy relies on iteratively changing the orientation of the solar sail once a certain threshold distance from the equilibrium point is surpassed. Thus, the satellite is always kept close enough to the non-eclipsed equilibrium point, allowing to maintain active communications. To do so, the dynamics of the system is linearized around the equilibrium point to study its behavior when the sail orientation changed. This allows us to determine the best sequence of orientation changes by a least squares approximation performed on each step. Finally, to test the robustness of the station-keeping strategy, several simulations for different values of the control parameters and the threshold distances are performed for a given period of mission.   
\section{Solar radiation pressure}
Solar radiation pressure (SRP) is the force generated by the impact of sunlight on a reflective surface. When we consider a solar sail, the SRP which mainly produces orbital perturbations, will be used as a method of spacecraft propulsion. 
The SRP exerted on the solar sail surface, $P$, due to the momentum carried by the Sun photons is given by\cite{mcinnessolar}: 
\begin{equation} \label{eq_presion}
P = \frac{W}{c},
\end{equation} 
where $c$ is the speed of light and $W$ the energy flux, i.e., the rate of transfer energy per unit area per unit time. The energy flux at a distance $r$ from the Sun can be written in terms of the solar luminosity $L_s$ and the distance from the Sun to the planet $R_{p}$ (referring to the planet, either Earth or Mars, with the subscript $p$) as:
\begin{equation} \label{eq_intermm1}
    W = W_{p} \left( \frac{R_{p}}{r} \right) ^2, \hspace{0.3 cm} \text{being} \hspace{0.2 cm} W_{p} = \frac{L_s}{4 \pi R_{p}^2},
\end{equation}
where $W_{p}$ is the energy flux measured at the planet's average distance from the Sun. Table \ref{tab:31} shows the parameters involved in the SRP computations for both the Earth-Sun and the Mars-Sun systems. 

\newcolumntype{g}{>{\columncolor{LightCyan}}c}
\begin{table}[ht]
\centering
\begin{tabular}{g| c| c| c} 
\hline
\rowcolor{Gray}
\text{Planet} &  $W$ (W$/$m$^2$ )&  $R_p$ (AU) & $P$ (N$/$m$^2$)  \\
\hline
Earth & 1367.56 & 1 & 4.5630 $\cdot 10^{-6}$  \\
Mars & 588.81 & 1.52 & 1.9646  $\cdot 10^{-6}$  \\ 
\hline
\end{tabular}
 \caption{SRP parameters for Earth and Mars}
 \label{tab:31}
\end{table}

\section{solar sail acceleration}
\begin{figure}[h!]
    \centering
    \includegraphics[scale = 0.45]{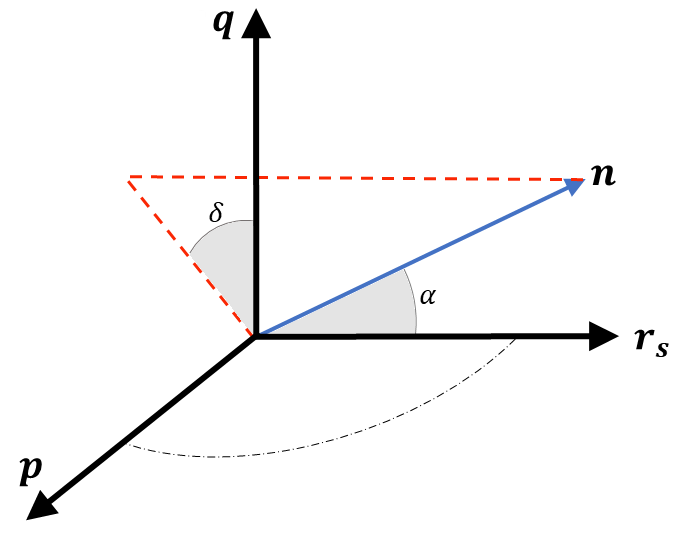}
    \caption{Solar sail orientation parameters in the reference frame  $\lbrace \vec{r_s}, \vec{p}, \vec{q} \rbrace $}
    \label{fig_vector_n_descompuesto}
\end{figure}

The force exerted by SRP causes an acceleration on a solar sail \cite{mcinnessolar}. This force depends on both, the area of the sail $A$ and its orientation relative to the Sun determined by the normal vector $\vec{n}$ to the solar sail surface. In this paper we assume a simple model for a solar sail perfectly reflecting, so the directions of the incident and reflected rays on the sail surface coincide, resulting on a force perpendicular to the sail. The incidence force, $\vec{F_i}$, is pointed towards the incident direction $\vec{u_i}$ while the reflected force $\vec{F_r}$ is in the opposite reflected direction $- \vec{u_r}$:
\begin{equation}  \label{eq_forces_refl_inc}
    \vec{F_i} = PA \langle \vec{u_i}, \vec{n} \rangle \vec{u_i}, \hspace{0.5 cm} \vec{F_r} = - P A \langle \vec{u_r}, \vec{n} \rangle \vec{u_r},
\end{equation}
where $P$ is the solar radiation pressure magnitude, $A \langle \vec{u_i}, \vec{n} \rangle$ the projection of the sail area in the $\vec{u_i}$ direction and $A \langle \vec{u_r}, \vec{n} \rangle$ in the $\vec{u_r}$ direction. 
\\\\
Since $\vec{u_i} - \vec{u_r} = 2 \langle \vec{u_i}, \vec{n} \rangle  \vec{n}$ and using \Eqref{eq_presion} and \Eqref{eq_intermm1}, the total force caused by the SRP, denoted as $ \vec{F_{SRP}}$, is given by: 
\begin{equation} \label{eq_fuerzaprevia}
    \vec{F_{SRP}}= \frac{2 A W_{p}}{c} \left( \frac{R_{p}}{r} \right) ^2 \langle \vec{u_i}, \vec{n} \rangle ^2 \vec{n} .
\end{equation}
The solar sail performance can be parameterized by a measure of a sail loading parameter $\sigma$. This parameter is defined as the total mass ratio per unit of sail area, $ \sigma = m/A$, playing a key role in the composition and design of the solar sail. Therefore, the total force \Eqref{eq_fuerzaprevia} exerted on the surface in terms of the sail loading is:
\begin{equation} \label{eq_fuerzatotal}
\vec{F_{SRP}} =  \frac{2W_{p}}{c} \frac{m}{\sigma} \left( \frac{R_{p}}{r} \right) ^2  \langle \vec{u_i} , \vec{n} \rangle ^2 \vec{n} .
\end{equation}
Considering a solar sail in a heliocentric orbit, the incidence direction $\vec{u_i}$ is set as the unit radial vector from the Sun to the sail $\vec{r_s}$. Then, the acceleration caused by SRP as a function of the gravitational acceleration experienced by the sail is expressed as:
\begin{equation} \label{aceleracion_grav}
    \vec{a_{SRP}}= \beta \frac{G M_s}{r^2 } \langle \vec{r_s}, \vec{n} \rangle ^2 \vec{n},
\end{equation}
where $G$ is the Universal Gravitational Constant, $M_s$ is the mass of the Sun and $\beta$ a dimensionless parameter known as the lightness number of the solar sail, which indicates the ratio between the SRP acceleration and the gravitational acceleration. The parameter $\beta$ is given by:
$$ \beta = \frac{\sigma^*}{\sigma}, 
\hspace{0.3 cm}  \text{being} \hspace{0.3 cm} 
\sigma^* = \frac{2 W R_{p} ^2}{ G M_s c} \approx 1.53 \hspace{0.1 cm} g/m^2.$$
As a consequence, taking into account that the solar radiation pressure acceleration is smaller than the gravitational attraction between the Sun and the sail, the $\beta$ values lies between $0$ and $1$. 
\\\\
To define the orientation of the solar sail force vector, let us start with the simplest case, considering a solar sail oriented perpendicular to the solar line (the line joining the Sun and the solar sail). In this case the normal vector $\vec{n}$ coincides with the  solar line vector $\vec{r_s}$, satisfying that $ \langle \vec{r_s}, \vec{n} \rangle = 1$. For an arbitrary orientation, the normal vector to the sail surface $\vec{n}$ can be parameterized by two angles, the sail cone angle $\alpha$ and the clock angle $\delta$. These angles indicate the orientation of the normal vector $\vec{n}$ with respect to the solar line vector $\vec{r_s}$. We consider an orthonormal reference frame whose origin is at the solar sail and a basis formed by $\lbrace \vec{r_s}, \vec{p}, \vec{q} \rbrace $ \cite{mcinnessolar}. Then, the normal vector $\vec{n}$ is expressed as:
\begin{equation} \label{def_vector_normal_n}
    \vec{n} = \cos \alpha \vec{r_s} + \sin \alpha \cos \delta \vec{p} + \sin \alpha \sin \delta \vec{q},
\end{equation}
where the $\vec{p}$ and $ \vec{q} $ vectors are:
\begin{equation*}
   \vec{p} = \frac{\vec{r_s} \times \vec{z}}{ | \vec{r_s} \times \vec{z}|}, \hspace{0.4 cm}  \vec{q} = \frac{(\vec{r_s} \times \vec{z}) \times \vec{r_s} }{| (\vec{r_s} \times \vec{z}) \times \vec{r_s} |},
\end{equation*}
being $\vec{z}$ the unit vector given by $\vec{z} = (0,0,1)$. Nevertheless, there are other possibilities to define these orientation angles\cite{farres2009contribution, mcinnessolar}.
\\\\
As a result, the sail cone angle $\alpha$ is defined as the angle between the sail normal vector $\vec{n}$ and the solar line vector $\vec{r_s}$, 
and takes values between $[-\pi/2, \pi/2]$ rad. While the sail clock angle $\delta$ is defined as the angle between the $\vec{q}$ reference direction and the projection of the sail normal vector $\vec{n}$ onto a plane orthogonal to the solar line vector $\vec{r_s}$, and takes values between  $[0, \pi]$ rad. In Figure \ref{fig_vector_n_descompuesto} the orientation angles which characterize the solar sail configuration are depicted.
\\
Note that since the normal vector $\vec{n}$ cannot point towards the Sun, then:
\begin{equation} \label{condiciones_existencia_soluciones1}
     \langle \vec{n}, \vec{r_s} \rangle = \cos \alpha  \geq 0.
\end{equation}
\subsection{Characteristic acceleration of a solar sail}
The solar sail efficiency can be also measured in terms of the characteristic acceleration, $a_0$, which is defined as the acceleration experienced by a perfectly reflecting solar sail oriented perpendicular to the solar line at a distance $1$ AU from the Sun:
$$ a_0 = \frac{2 P A  \rho_r}{m},$$ 
where $\rho_r$ is the reflectivity coefficient of the sail, assumed to be $1$.
Tables \ref{tab:3} and \ref{tab:4} present some relationships between the solar sail lightness number $\beta$, the sail loading parameter $\sigma$ and the characteristic acceleration $a_0$ for the Earth-Sun and Mars-Sun systems respectively. It should be noted that the characteristic acceleration $a_0$ from Table \ref{tab:4} is computed at a distance $1.52$ AU for the Mars-Sun case. In this paper, we consider a $3$U CubeSat nanosatellite, corresponding to a mass of $4.5$ kg, equipped with a perfectly reflecting solar sail.
\begin{table*}
           \centering
           \captionsetup[subtable]{position = below}
          \captionsetup[table]{position=top}
         \caption{Relation between $\beta$, $\sigma$, $A$ and $a_0$ }
           \begin{subtable}{0.35\linewidth}
               \centering
               \begin{tabular}{|c| c| c| c|} 
\hline 
 $\beta$ & $\sigma$ (g/m$^2$) & A (m$^2$) & $a_0$ (mm/s$^2$)  \\ 
 \hline \hline
 \rowcolor{Gray}
0.0102 & 150 &  30 & 0.06113\\
 0.0204 & 75 &  60& 0.12226  \\ 
 \rowcolor{Gray}
 0.0282 & 54.21  & 83 & 0.16913\\
  0.0506 & 30.2   & 149 & 0.3036\\
  \rowcolor{Gray}
0.102 & 15  & 300 & 0.6113\\
0.306 & 5  & 900 & 1.834\\
\hline
\end{tabular}
               \caption{Earth-Sun system}
               \label{tab:3}
           \end{subtable}%
           \hspace*{4em}
           \begin{subtable}{0.65\linewidth}
               \centering
          \begin{tabular}{|c| c| c| c|} 
\hline 
 $\beta$ & $\sigma$ (g/m$^2$)  & A (m$^2$) & $a_0$ (mm/s$^2$) \\ 
 \hline \hline
 \rowcolor{Gray}
0.0102 & 150  & 30 & 0.0262\\
 0.0205 & 75  & 60 & 0.0524 \\ 
 \rowcolor{Gray}
  0.0283 & 54.21 & 83   & 0.0724\\
 0.0509 & 30.2  & 149 & 0.13 \\
 \rowcolor{Gray}
0.1025 & 15  & 300 & 0.2618\\
0.307 & 5  & 900 & 0.7855\\
\hline
\end{tabular} 
                \caption{Mars-Sun system}
                 \label{tab:4}
           \end{subtable}
       \end{table*}\\\\
Comparing both Tables \ref{tab:3} and \ref{tab:4}, we can see that for similar $\beta$ values, the characteristic accelerations of a sail in the Mars-Sun system are smaller than in the Earth-Sun system. 
\noindent Notice that Tables \ref{tab:3} and \ref{tab:4} only include solar sails with areas up to $900$ m$^2$. No greater values have been considered, as the largest sail ever designed had a total area of approximately $1600$ m$^2$\footnote{\url{https://www.planetary.org/space-missions/solar-cruiser}}. Moreover, areas smaller than $30$ m$^2$ would correspond to low thrust propulsion practically of the order of 10$^{-4}$ N and therefore have not been included in this study.
\section{Dynamical model for a solar sail}
\begin{figure}[h!]
\centering
\includegraphics[scale = 0.5]{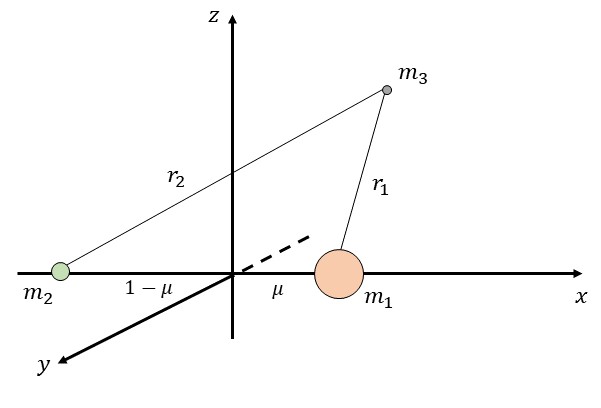}
\caption{Representation of the three bodies in the rotating reference frame. (Not to scale)}
\label{fig:rotatorio}
\end{figure}

A suitable model to describe the dynamics of a satellite equipped with a solar sail in the vicinity of Earth or Mars is the circular Restricted Three Body Problem (RTBP)\cite{Valtonen} including the effect of solar radiation pressure. 
\\\\
This model considers two primary bodies, $m_1$ and $m_2$, (by convention $m_1 > m_2$) and a small third body with mass $m_3$ assumed to be negligible compared to the other two. All bodies are supposed to be spherically symmetric, considered as point masses, with the motion of the two primaries contained in the same plane. The dynamics of the bodies is governed by the Law of Universal Gravitation, assuming that $ m_1$ and $m_2$ describe circular orbits around the center of mass of both bodies, and $m_3$ does not affect the motion of the two primary masses. Hence, the equations of motion of the third body (corresponding to the satellite equipped with the solar sail), subject only to the gravitational attraction as well as the acceleration due to SRP are obtained. 
\\\\
Let us consider a rotating reference frame $(x,y,z)$ with the origin at the center of mass of the two primary bodies, the $x$ axis is taken in the direction joining the two primaries, the $z$ axis is perpendicular to the ecliptic plane and the $y$ axis is chosen to be oriented positively and orthogonal to the other two axes. To simplify, the system is expressed with normalized units of mass, distance and time, such that the total mass of the system is $1$, the distance between the primaries is $1$ and their orbital period about the center of mass is $2 \pi$. The Universal Gravitational Constant then becomes $G = 1$. With this normalized units, the mass of the big primary body is $1 - \mu$, located at $(\mu,0,0)$, while the mass of the small primary body is $\mu$, located at $(-(1 - \mu),0,0)$, where $\mu \in [0, \frac{1}{2}]$ is referred as the mass parameter, given by $\mu = m_2/(m_1 + m_2)$\cite{SZEBEHELY19677}. The values of the mass parameters for both the Earth-Sun and the Mars-Sun systems are respectively $\mu_{_{ES}} = 3.003480 \cdot 10^{-6}$  and $\mu_{_{MS}} = 3.227155 \cdot 10^{-7}$. Furthermore, the distances from the third body to the primary masses are denoted by $r_1$ and $r_2$. 
In Figure \ref{fig:rotatorio}, a scheme of the RTBP model in the dimensionless units is depicted. 
\\
In this rotating coordinate system, the equations of motion for the third body can be obtained using the variational principle of least action, described by the Euler-Lagrange equations:
\begin{equation} \label{eulerlagrangerotatorio}
   \frac{d}{dt} \frac{\partial \mathcal{L}  }{\partial\dot{q}^i} - \frac{\partial  \mathcal{L} }{\partial q^i } = 0, 
\end{equation}
where $q^i$ are the generalized coordinates ---in this case $x,y,z$--- and $\mathcal{L}$ the Lagrangian of the system, expressed as $  \mathcal{L}  = T - U$, being $T$ and $U$ the kinetic and potential energy respectively. Moreover, the acceleration $\vec{a_{SRP}} $ caused by the SRP effect in this normalized rotating reference frame is:
\begin{equation} \label{aceleracion_SRP_defff}
     \vec{a_{SRP}} = \beta \frac{1 - \mu}{r_1^2 }  \langle \vec{r_s }, \vec{n} \rangle ^2 \vec{n}, \hspace{0.3 cm} \text{where} \hspace{0.2 cm}   \vec{r_s} = \frac{(x - \mu, y , z) }{r_1}.
\end{equation}
Thus, under the gravitational attraction of the two primary bodies, the system of non-linear ordinary differential equations of motion described by a satellite equipped with a solar sail --- parameterized by the two angles of orientation $\alpha$ and $\delta$---, are:
\begin{equation} \label{ecuacion_catalogue}
    \begin{cases}
    \vspace{0.09 cm}
     \ddot{x} - 2 \dot{y} = \frac{\partial \bar{U}}{\partial x} + \beta \frac{1 - \mu}{r_1^2} \cos^2 \alpha \left( \frac{ -(x - \mu)z}{s r_1} \sin \alpha \cos \delta + \frac{y}{s} \sin \alpha \sin \delta \right),\\ 
      \ddot{y} + 2 \dot{x} = \frac{\partial \bar{U}}{\partial y} + \beta \frac{1 - \mu}{r_1^2} \cos^2 \alpha \left( \frac{ -yz}{s r_1} \sin \alpha \cos \delta - \frac{x - \mu}{s} \sin \alpha \sin \delta \right),\\
\ddot{z} = \frac{\partial \bar{U}}{\partial z} + \beta \frac{1 - \mu}{r_1^2} \cos^2 \alpha \left( \frac{ s}{ r_1} \sin \alpha \cos \delta \right),
    \end{cases}
\end{equation}
where $s^2 = (x - \mu)^2 + y^2$ and the effective potential $ \bar{U}$, including the effect of a solar sail, is written as:
\begin{equation} \label{potencial_efectivo_definitivo}
    \bar{U}(x,y,z) = \frac{1}{2} (x^2 + y^2) + (1 - \beta \cos^3 \alpha) \frac{(1 - \mu) }{r_1} + \frac{\mu}{r_2}, 
\end{equation}
being $r_1^2 = (x - \mu)^2 + y^2 + z^2$ and $r_2^2 = (x + 1 - \mu)^2 + y^2 + z^2$ the distances from the sail to the first and second body respectively.
\\\\
Due to the complexity of this system of non-linear second order equations of motion \Eqref{ecuacion_catalogue}, 
a qualitative study of the behavior of the solutions is carried out with special emphasis on the determination and classification of the equilibrium points for the separate Earth-Sun and Mars-Sun systems and the study of the linearized dynamics.
\section{Equilibrium points of the equations of motion}
In this section, we describe the process followed to obtain the equilibrium points of the system by distinguishing between two cases of solar sail orientation: (a) an orientation perpendicular to the solar line, and (b) an arbitrary orientation. 
\subsection{Solar sail oriented perpendicular to the solar line}
This is the simplest case, when the solar sail is oriented perpendicular to the solar line, the normal vector to the sail surface $\vec{n}$ and the unit vector in the solar line direction $\vec{r_s}$ are coincident, verifying $ \alpha = \delta = 0$. To obtain the equilibrium points, we seek for the stationary solutions of the equations of motion \Eqref{ecuacion_catalogue}, i.e. $\dot{x} = \dot{y} = \dot{z} = \ddot{x} = \ddot{y} = \ddot{z} = 0$, where the solutions correspond to the critical points of the effective potential \Eqref{potencial_efectivo_definitivo}. This effective potential has 5 critical points, corresponding to the 5 equilibrium points parameterized by $\beta$, known as Sub$-L_i$ for $i \in \lbrace 1,\dots,5 \rbrace$ and denoted by $SL_i$. Three of these equilibrium points $SL_{1,2,3}$ are obtained by solving three quintic equations, which can be numerically determined by Newton's method, in the form:
$$SL_1 = (\mu -1 + \xi_1, 0, 0), \hspace{0.4 cm} SL_2 = (\mu -1 - \xi_2, 0, 0), \hspace{0.4 cm} SL_3 = (\mu + \xi_3, 0, 0),$$
where $\xi_1, \xi_2$ and $\xi_3$ are the admissible roots in the valid intervals for each of the quintic equations\cite{farres2009contribution}. While the coordinates of the two other equilibrium points, $SL_{4}$ and $SL_{5}$, are respectively given by:
$$    x = \mu - \frac{(1 - \beta)^{2/3}}{2}, \hspace{0.4 cm}     y = \pm (1 - \beta)^{1/3} \left[ 1 - \frac{(1 - \beta)^{2/3}}{4} \right]^{1/2}.$$
$SL_{1,2,3}$ are collinear points located on the line joining the two primary bodies for $ y = z = 0$ while $SL_{4,5}$ are positioned at the apexes of an isosceles triangle formed with the primaries for $y \neq 0$. Note that when $\beta$ tends to $1$, the equilibrium point $SL_2$ approaches to the planet (either Earth or Mars) while the other equilibrium points get closer to the Sun. 
\\\\
For the classical RTBP, when the SRP is neglected (i.e. considering $\beta = 0$ in \Eqref{ecuacion_catalogue}), five equilibrium points are also obtained, denoted by $L_i$ for $ i \in \lbrace 1, \dots, 5 \rbrace$, known as Lagrange points or libration points. The geometrical distribution of these equilibrium points is similar to the one described above\cite{koon2000dynamical}. 
\subsection{Solar sail arbitrarily  oriented to the solar line}
\begin{figure}[h!]
\begin{center}
    \textbf{Earth-Sun system} ($\delta = \pi/2$, $\alpha \in [- \pi/2,  \pi/2] $) \par\medskip
    \includegraphics[scale=0.33]{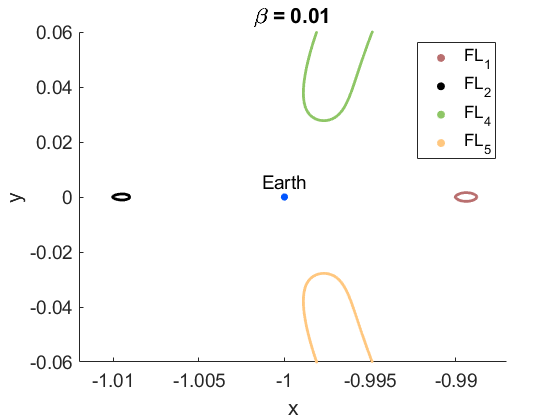}\hspace{0.1 cm}
    \includegraphics[scale=0.33]{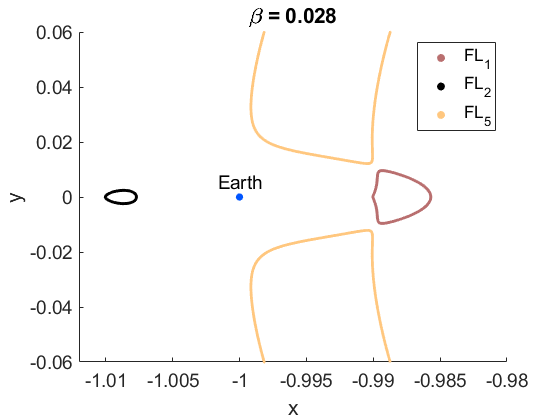}\hspace{0.1 cm}
    \includegraphics[scale=0.33]{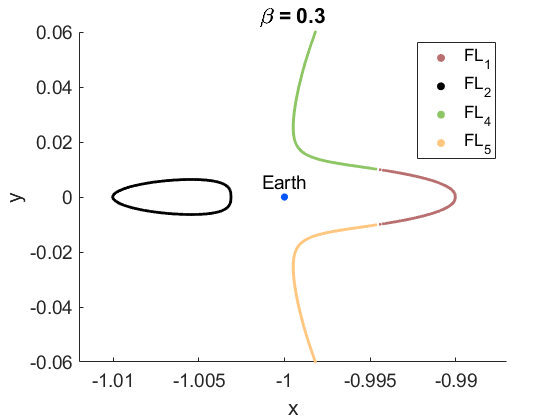}\hspace{0.1 cm}
        \\[\smallskipamount]
            \textbf{Mars-Sun system} ($\delta = \pi/2$, $\alpha \in [- \pi/2,  \pi/2] $) \par\medskip
        \includegraphics[scale=0.33]{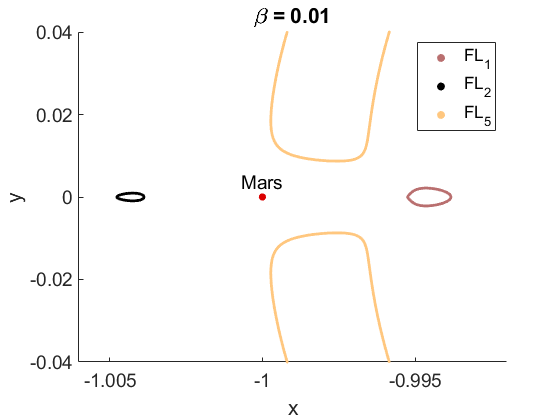}\hspace{0.1 cm}
    \includegraphics[scale=0.33]{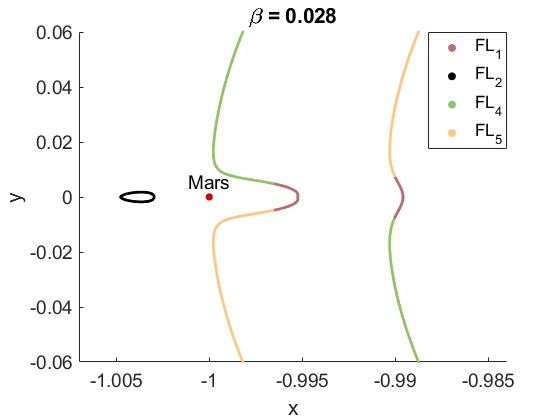}\hspace{0.1 cm}
    \includegraphics[scale=0.33]{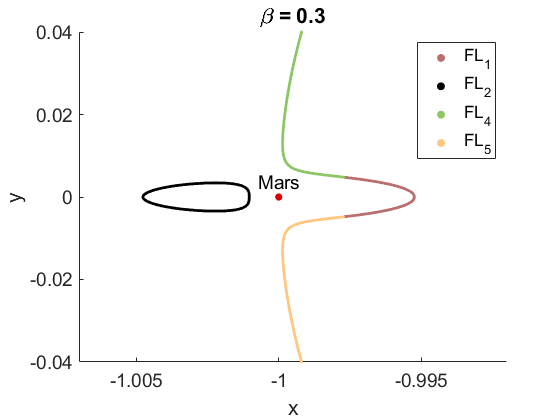}\hspace{0.1 cm}
    \caption{Families of equilibrium points centered on planets for \textit{xy} plane}
    \label{fig_ptos_all_TierrayMarte}
    \end{center}
\end{figure}

\begin{figure}[h!]
\begin{center}
    \textbf{Earth-Sun system} ($\delta = 0$, $\alpha \in [- \pi/2,  \pi/2] $) \par\medskip
    \includegraphics[scale=0.33]{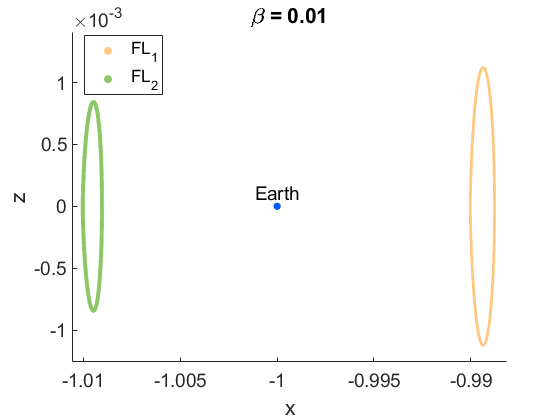}\hspace{0.1 cm}
    \includegraphics[scale=0.33]{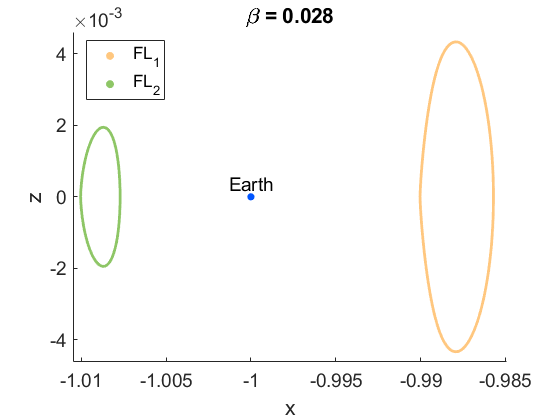}\hspace{0.08 cm}
    \includegraphics[scale=0.33]{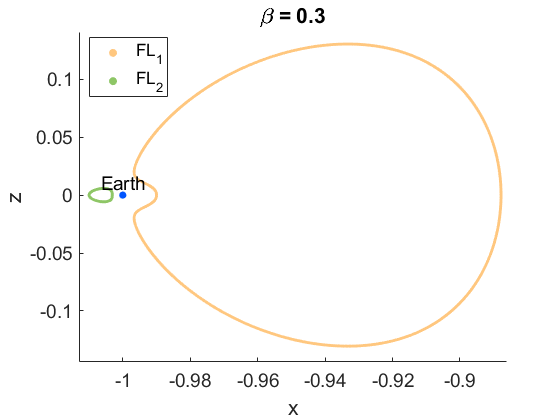}\hspace{0.1 cm}
        \\[\smallskipamount]
            \textbf{Mars-Sun system} ($\delta = 0$, $\alpha \in [- \pi/2,  \pi/2] $)\par\medskip
        \includegraphics[scale=0.33]{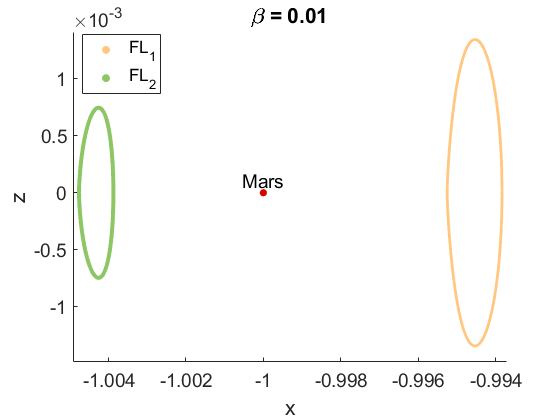}\hspace{0.1 cm}
    \includegraphics[scale=0.33]{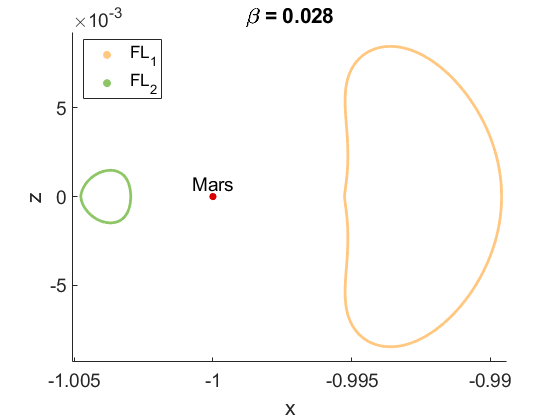}\hspace{0.15 cm}
    \includegraphics[scale=0.33]{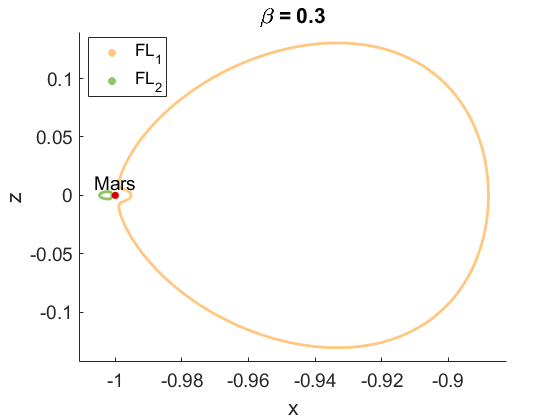}\hspace{0.04 cm}
    \caption{Families of equilibrium points centered on planets for \textit{xz} plane}
    \label{fig_ptos_all_TierrayMarte_planoxz}
    \end{center}
\end{figure}

\begin{figure}[h!]
\centering
\begin{subfigure}{0.27\textwidth}
  \centering
  \includegraphics[scale = 0.28]{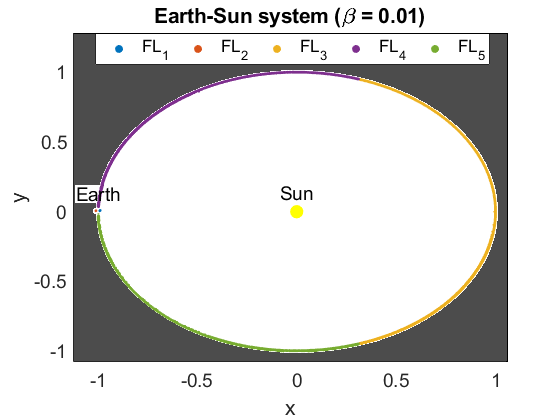}
\end{subfigure}
\begin{subfigure}{0.26\textwidth}
  \centering
  \includegraphics[scale = 0.28]{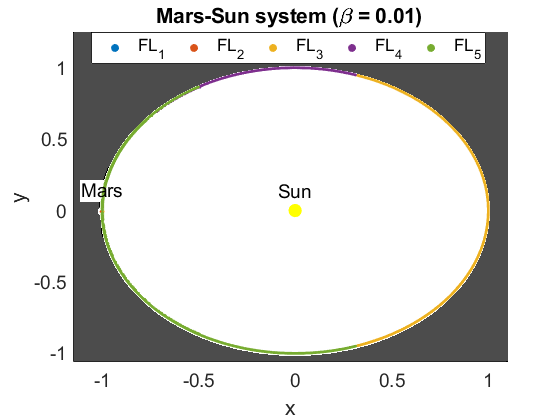}
\end{subfigure}
\begin{subfigure}{0.23\textwidth}
  \includegraphics[scale = 0.28]{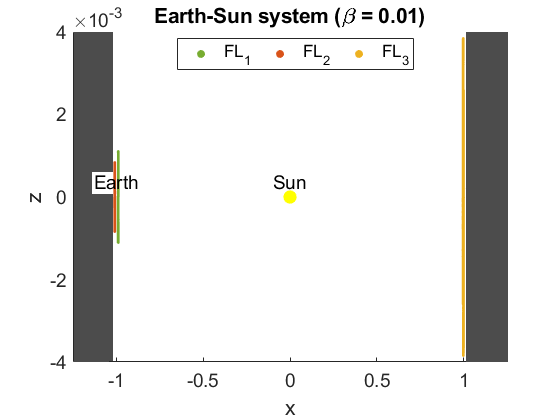}
\end{subfigure}
\begin{subfigure}{0.22\textwidth}
  \centering
  \includegraphics[scale = 0.28]{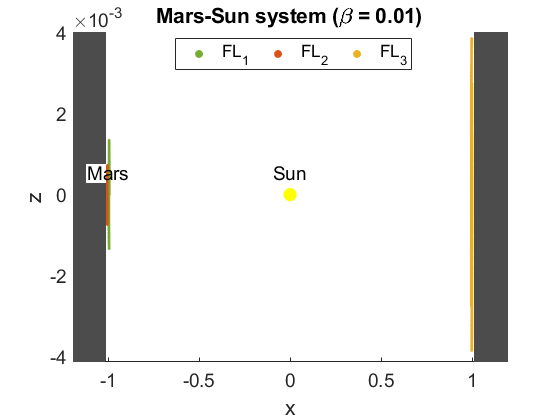}
\end{subfigure}
\caption{Non-existence regions of equilibrium points}
\label{sistema_planteta_Sun_001}
\end{figure}

\begin{figure}[h!]
\begin{center}
    \textbf{Earth-Sun system} ($\delta = \pi/2$, $\alpha \in [- \pi/2,  \pi/2] $) \par\medskip
    \includegraphics[scale=0.33]{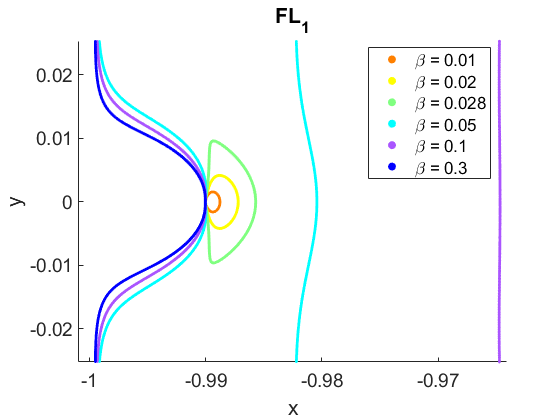}\hspace{0.1 cm}
    \includegraphics[scale=0.33]{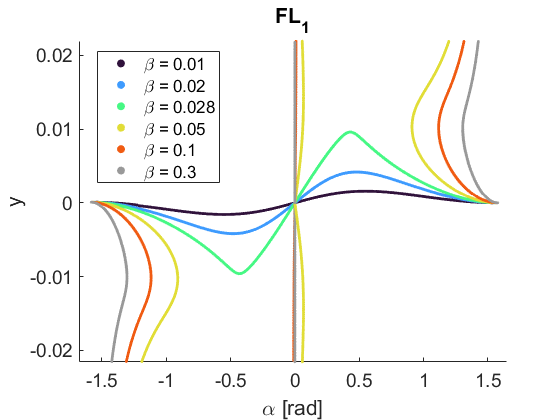}\hspace{0.1 cm}
        \includegraphics[scale=0.22]{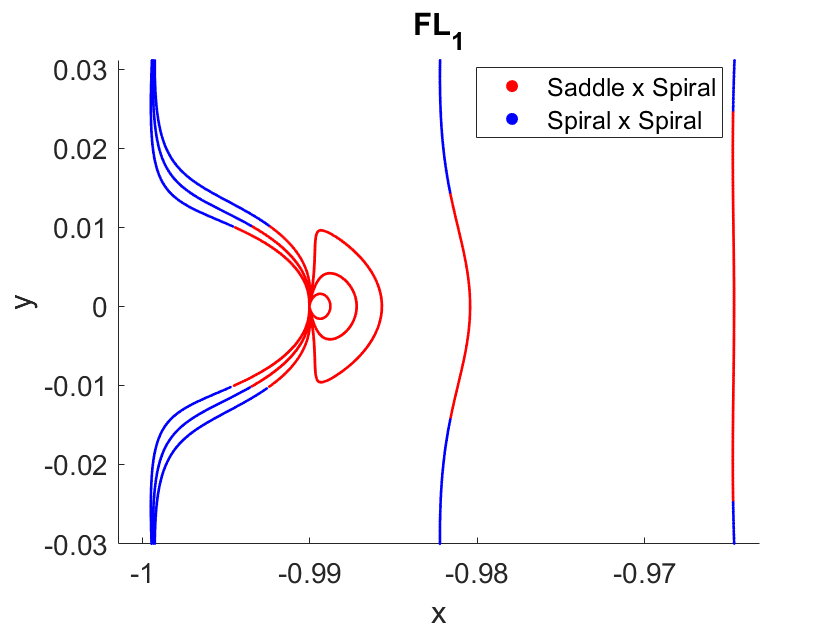}\hspace{0.1 cm}
        \\[\smallskipamount]
            \textbf{Mars-Sun system} ($\delta = \pi/2$, $\alpha \in [- \pi/2,  \pi/2] $)\par\medskip
        \includegraphics[scale=0.33]{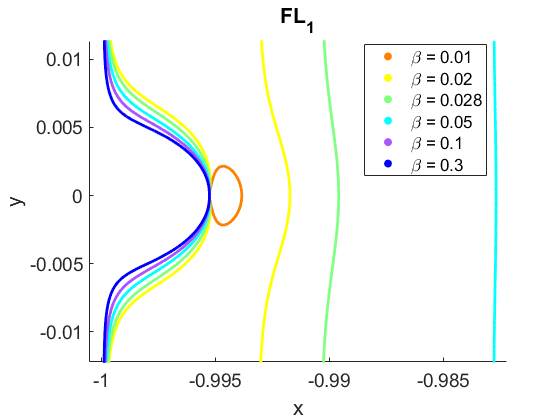}\hspace{0.1 cm}
    \includegraphics[scale=0.33]{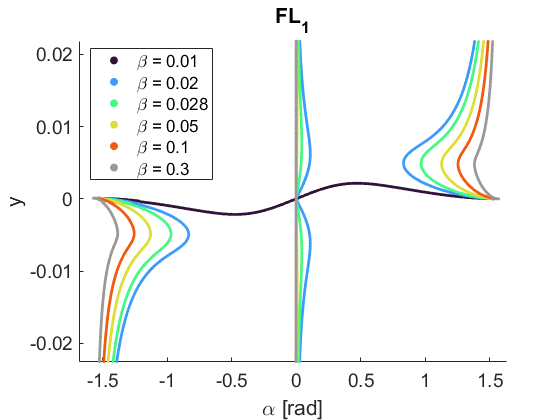}\hspace{0.1 cm}
        \includegraphics[scale=0.33]{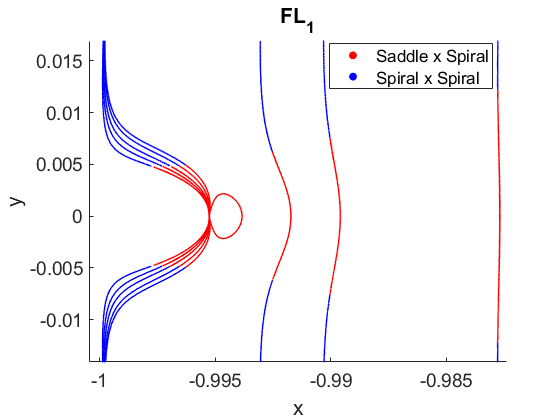}\hspace{0.1 cm}
    \caption{Families of equilibrium points centered on planets for \textit{xy} plane}
    \label{fig_familia_FL1_solamente_XY}
    \end{center}
\end{figure}

\begin{figure}[h!]
\begin{center}
    \textbf{Earth-Sun system} ($\delta = 0$, $\alpha \in [- \pi/2,  \pi/2] $) \par\medskip
    \includegraphics[scale=0.33]{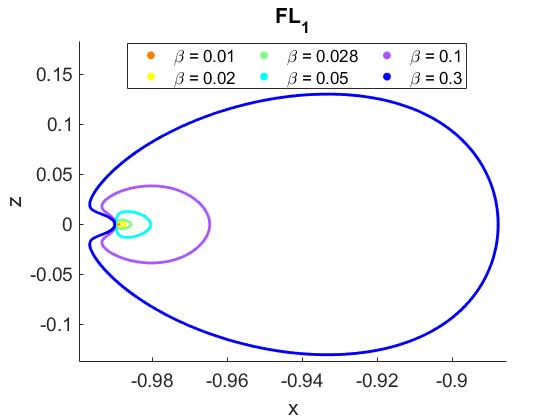}\hspace{0.1 cm}
    \includegraphics[scale=0.33]{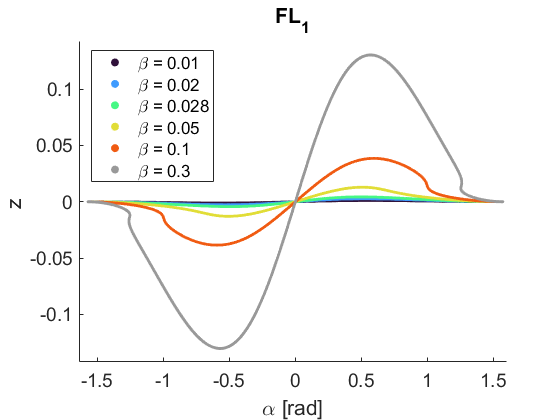}\hspace{0.13 cm}
    \includegraphics[scale=0.33]{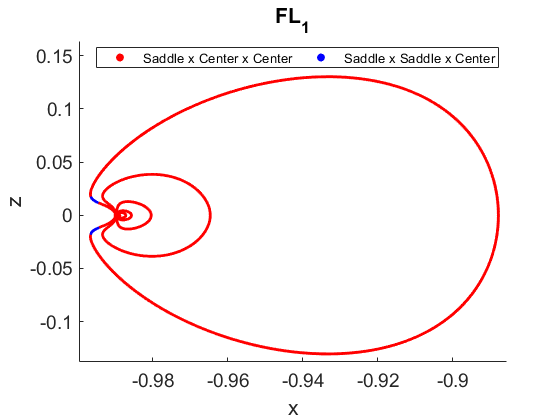}\hspace{0.13 cm}
        \\[\smallskipamount]
            \textbf{Mars-Sun system} ($\delta = 0$, $\alpha \in [- \pi/2,  \pi/2] $)\par\medskip
        \includegraphics[scale=0.33]{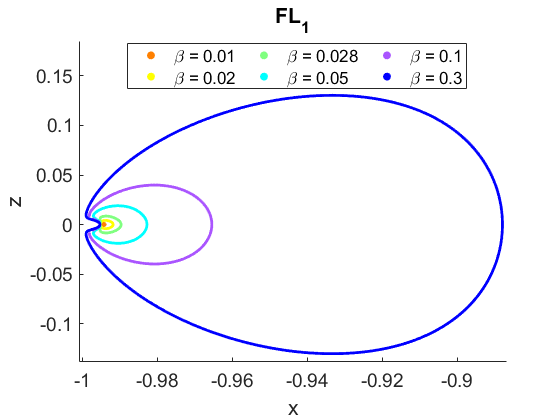}\hspace{0.1 cm}
    \includegraphics[scale=0.33]{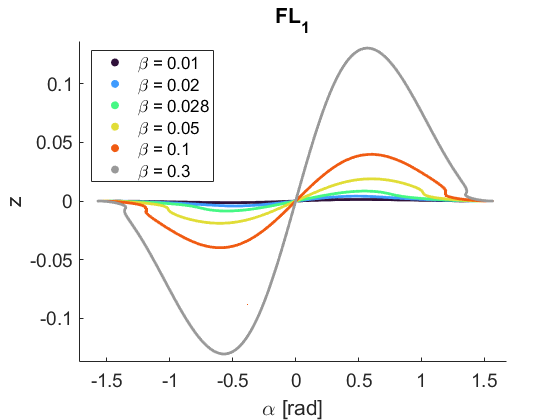}\hspace{0.1 cm}
    \includegraphics[scale=0.33]{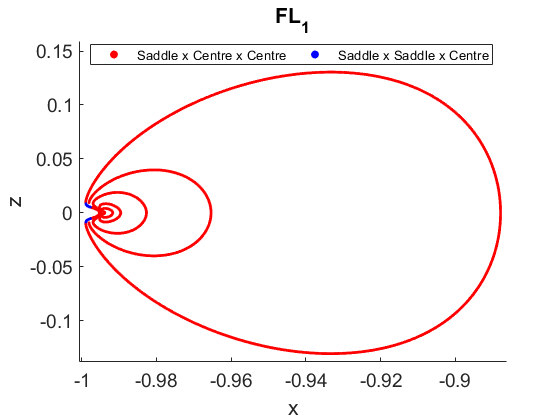}\hspace{0.13 cm}
    \caption{Families of equilibrium points centered on planets for \textit{xz} plane}
    \label{fig_familia_FL1_solamente_XZ}
    \end{center}
\end{figure}

When the solar sail has an arbitrary orientation to the solar line, (i.e. $\alpha \neq 0$), the equations of motion \Eqref{ecuacion_catalogue} are parameterized by the two orientation angles $\alpha$ and $\delta$. As before, the equilibrium points are the $(x, y, z)$ solutions of equating the right-hand side of \Eqref{ecuacion_catalogue} to zero. However, no explicit formula can be obtained, so a numerical method is used to determine the position of these equilibrium points for different sail orientation angles $\alpha$ and $\delta$. Even though there are several methods to obtain the equilibrium points\cite{Aliasi, McInnes_Colin_Alastair, McInnes_Colin_otro}, we use a continuation method\cite{Krauskopf} by considering a fixed value of the clock angle $\delta$ and varying the cone angle $\alpha$ in its domain $[-\frac{\pi}{2},\frac{\pi}{2}]$. The continuation method used to find the location of these equilibrium points is described in the next section.
\\\\
In this paper, two possible case studies have been considered. Firstly, we consider a fixed value of $\delta = \pi/2$ and varying $\alpha$, where the equilibrium points are displacing in the ecliptic plane $z = 0$, (as $a_{SRP_z} = 0$). Secondly, we take $\delta = 0$ and also varying $\alpha$, now the equilibrium points lie above or below the ecliptic plane, with $y = 0$, (as $a_{SRP_y} = 0$).
\\\\
In the $xy$ plane 5 continuous families of equilibrium points parameterized by $\alpha$, denoted by $FL_i$ for $ i \in \lbrace 1, \dots, 5 \rbrace$, are obtained, while in the $xz$ plane, there are only 3 families of equilibrium points  $FL_{1,2,3}$. These uniparametric families of equilibrium points are also referred to \textit{artificial} equilibria as we can think of them as artificially displacing the location of $SL_i$ for different values of the sail orientation. Note that each of these families contains their respective Lagrange point $L_i$ and also the corresponding $SL_i$ for each $ i \in \lbrace 1, \dots, 5 \rbrace$. In Figures \ref{fig_ptos_all_TierrayMarte} and \ref{fig_ptos_all_TierrayMarte_planoxz}, the families of equilibrium points are respectively depicted for certain values of $\beta = 0.01, 0.028$ and $0.3$ in the $xy$ plane and in the $xz$ plane, for Earth (top plots) and Mars (bottom plots). 
\\\\
In the $xy$ plane, as $\beta$ increases, the families of equilibrium points are merged with each other, thus generating connected sets. In the $xz$ plane, these continuous families are non-connected sets. For larger values of $\beta$, these families grow up, but do not merge into each other. Note that $FL_3$ family is not represented since it is located farther away. 
\\\\
The shape described by these families of equilibrium points may be explained from the condition in \Eqref{condiciones_existencia_soluciones1} for the limitation of the solar sail orientation \cite{mcinnessolar}. When the normal vector $\vec{n}$ does not have the same direction as the solar line vector $\vec{r_s}$, i.e. $ \langle \vec{r_s}, - \Delta \Bar{U} \rangle \geq 0$, then yielding:
\begin{equation} \label{region_S_ptos_eq}
    S(x,y,z) := x (x - \mu) + y^2 - \frac{1 - \mu }{r_1} -  \frac{\mu \langle \vec{r_1},\vec{r_2} \rangle }{r_2^3} = 0,
\end{equation}
which defines the boundary for the existence of equilibrium points. This function \eqref{region_S_ptos_eq} has two disconnected regions $S_1$ and $S_2$ that determine the required boundary. The $S_1$ region is defined by the boundaries $x < FL_2$ and $ x> FL_3$, while the $S_2$ region is delimited by $-1 + \mu < x < FL_1$. Figures \ref{sistema_planteta_Sun_001} shows the location of the equilibrium points for a given value of $\beta = 0.01$ as well as the non-existence regions of equilibria, $S_1$ and $S_2$, shaded in gray, for each dimensionless Earth-Sun and Mars-Sun system in the $z = 0$ and $y = 0$ planes.
 \\
From now on, due to the location and geometry of these families of equilibrium points on the $xy$ and $xz$ planes, we reduce our study to the family of equilibrium points $FL_1$, as this family of equilibrium points is the closest to the respective planet (Earth or Mars) in each of the systems, as well as the closest to the Sun. Moreover, its location is advantageous for the Earth-Mars communication mission. 
\\\\
The stability and classification of the points in the $FL_1$ family of equilibria are also discussed. The stability of these equilibrium points is described by the eigenvalues of the Jacobian matrix around each equilibrium point, where we have that $FL_1$ is unstable both in the $xy$ and $xz$ planes, having in all cases at least one eigenvalue with positive real part. Moreover, due to the fact that the dynamical system preserves areas, the sum of all the eigenvalues is zero. Then, the equilibrium points of this family $FL_1$ are classified by pairs of eigenvalues on each of the planes, using the symbol $\times$ to distinguish pairwise groupings. 
\\\\
For $\delta = \pi/2$ , all the families of equilibrium points are in an invariant manifold $ \lbrace z = \dot{z} = \Ddot{z} = 0\rbrace $. As a result, the dynamics of the family $FL_1$ in the $xy$ plane is of the type saddle $\times$ spiral (a pair of opposite real eigenvalues and another pair of conjugate complex with non-zero real part) and spiral  $\times$ spiral (two pair of conjugate complex eigenvalues with non-zero real part). While for $\delta = 0$, the plane $y=0$ is not invariant and we cannot reduce the phase space dimension. Then, the family $FL_1$ contains 6-dimensional equilibrium points of the type saddle $\times$ center  $\times$ center (a pair of opposite real eigenvalues, denoted by $ \pm \gamma$, with $\gamma > 0$, and two pairs of conjugate complex ones with zero real part, denoted by $\pm \eta i$ and $\pm \nu i$) and saddle $\times$ saddle  $\times$ center (two pair of opposite real eigenvalues and pair of conjugate complex eigenvalues with zero real part). Figures \ref{fig_familia_FL1_solamente_XY} and \ref{fig_familia_FL1_solamente_XZ} show, for several values of $\beta$, the family of equilibrium points $FL_1$ (left-side) and the classification of these equilibria of the $FL_1$ family (right-side) based on its linear dynamics in the dimensionless Earth-Sun and Mars-Sun systems for the $xy$ and $xz$ planes respectively. Note that the classification of these equilibria is reflected using different colors. Moreover, the variation in $y$ and $z$ coordinates of the artificial equilibrium points as a function of the sail orientation angle $\alpha$ is also plotted (center-side).  
\\
\\
Notice that in the $xy$ plane, as the values of $\beta$ increase, the continuation method describes a non-connected set $FL_1$ family of equilibria. This behavior is reflected in plots of the coordinate $y$ as a function of $\alpha$, where symmetries are presented with respect to the orientation angle $\alpha = 0$. Note that for larger values of $\beta$, the range of admissible sail orientations $\alpha$ decreases. As a result, at certain sail orientations we do not have equilibrium points in the $FL_1$ family. On the other hand, in the $xz$ plane, the connected set $FL_1$ has equilibrium points for all values of $\alpha$. Moreover, the growth of this family of equilibrium points is notably higher compared to the $xy$ plane, thus obtaining much larger sets of equilibrium points. The Lagrange points $L_i$ and $SL_i$ for $ i \in \lbrace 1, \dots, 5 \rbrace$, are also unstable equilibrium points\cite{koon2000dynamical}. The linear dynamics for $L_{1,2,3}$ and $SL_{1,2,3}$ is saddle $\times$ center $\times$ center while for $L_{4,5}$ and $SL_{4,5}$ is center $\times$ center $\times$ center.
\section{Continuation method}  
As we have described, the equilibrium points do not have an explicit equation for their position as a function of $\beta$, $\alpha$, $\delta$. Hence we use a continuation method to derive their location. Let us briefly describe the continuation method that we have used in this problem. 
\\\\
For a fixed value of $\delta$, the determination of the equilibrium points consists of solving for each $\alpha$, the nonlinear system of algebraic equations parameterized by $\alpha$ that results of equating the right-hand side of \Eqref{ecuacion_catalogue} to zero. In a more general framework, they can be expressed implicitly as:
\begin{equation} \label{function_F}
    F(x, \alpha) = 0, \hspace{0.3 cm} \text{with} \hspace{0.1 cm} F: \mathbb{R}^2 \times \mathbb{R} \to \mathbb{R}^2 \hspace{0.1 cm}  \text{regular.}
\end{equation}
Notice that for the particular value $\alpha_0 = - \pi/2$, the equilibrium points of \Eqref{function_F} correspond to those of the RTBP, as the effect of the solar sail is canceled. Thus, for this $\alpha_0$ the solutions correspond to the Lagrange points $L_i$ for $ i \in \lbrace 1, \dots, 5 \rbrace$. Consequently, for $\alpha_1 = \alpha_0 + h_0$, with a certain spatial step size $0 < h_0 \ll 1$, each Lagrange point can be employed as an initial estimate of the solutions of $F(x, \alpha_1) = 0$. These solutions can be computed iteratively using Newton's method as:
\begin{equation} \label{iteraciones_Newton}
\begin{cases} 
    x_{k + 1} = x_k - (JF(x_k, \alpha_1))^{-1} F(x_k, \alpha_1), \hspace{0.3 cm} k = \lbrace 0,1,2, \dots \rbrace \\
     x_0 = L_i \quad \text{for each $i$,}
\end{cases}
\end{equation}
where $JF$ is the Jacobian matrix of $F$ with respect to $x$. For efficiency and numerical stability we avoid calculating the inverse of the Jacobian \Eqref{iteraciones_Newton} by solving the linear system given by:
\begin{equation}
    \label{newton2}
    JF(x_k,\alpha_1) \cdot h = - F(x_k, \alpha_1), \hspace{0.2 cm} \text{with} \hspace{0.1 cm} x_{k+1} = h + x_k. 
\end{equation}
Then, for a set tolerance $\epsilon$, the method ends when $\norm{h}< \epsilon$. Note that in general, for an arbitrary $F$, if the system is sufficiently complex, it is not easy to have a good initial estimate of the zeros of \Eqref{function_F} and a random or badly chosen initial value may cause the iterations of Newton's method \Eqref{iteraciones_Newton} to not converge properly to any solution. However, as we have just mentioned, in this case for $\alpha_0 = - \pi/2$ we precisely know the zeros of $ F(x, \alpha_0) = 0$ (corresponding to the Lagrange points) and we can use this as an initial guess for $\alpha_1$ = $\alpha_0 + h_0$ (provided $h_0$ is small).
\\\\
In the iterative process, a non-uniform discretization of the interval $[- \pi/2, \pi/2]$ is considered for computational efficiency. This discretization is formed by a set of $M + 1$ values of $\alpha$, this is $-\frac{\pi}{2} = \alpha_0 < \alpha_1 < \dots < \alpha_M = \frac{\pi}{2}$, so the spatial step size between $\alpha_n$ and $\alpha_{n+1}$ is not necessarily the same for all  $ n \in \lbrace 0, \dots, M-1 \rbrace$. We denote this step size by $h_n > 0$, which is determined by the convergence rate of the Newton's method for $\alpha_n$. It will in fact be on this set of values $\alpha_i$ for which we will iteratively determine the equilibrium points from the known solution for $\alpha_{i -1}$. Note that this discretization may save unnecessary calculations. Particularly, on smoother points of the solution curve parameterized by $\alpha$, a greater step may be sufficient to ensure a fast enough convergence of Newton’s method. However, at points with larger irregularities, smaller steps are required.
\\

\begin{figure}[h!]
    \centering
     \includegraphics[scale = 0.27]{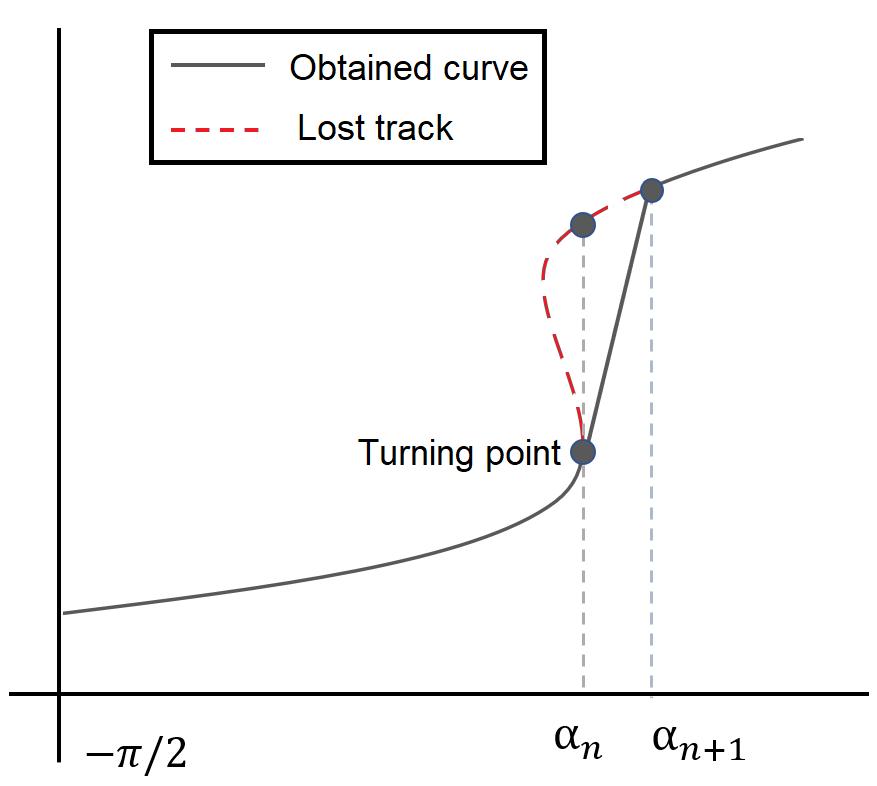}
     \caption{Turning point in the solution curve}
  \label{fig_return_point}
\end{figure}

\begin{figure}[h!]
    \centering
\includegraphics[scale = 0.18]{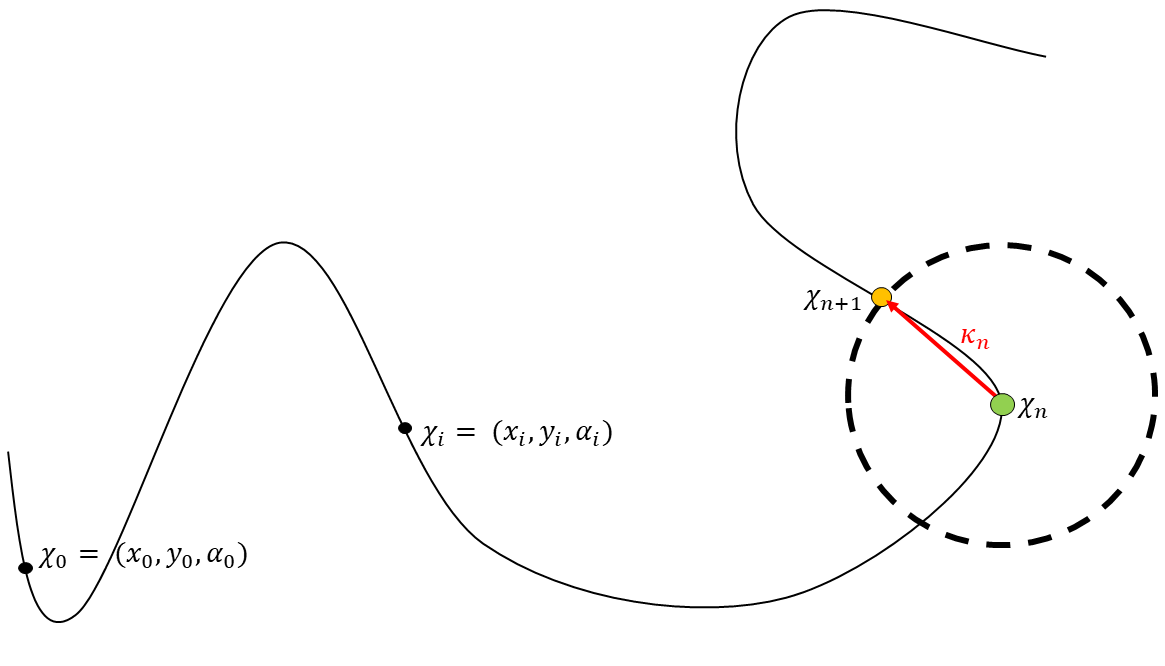}
     \caption{Proposed solution in the presence of a turning point}
  \label{fig_return_point2}
\end{figure}

\noindent These uniparametric solution curves have been obtained by continuously increasing the values of $\alpha$ in the continuation method. However, a problem may arise due to the existence of the so-called \textit{turning points}. For a certain $\alpha$, a turning point is a point at which the geometry of the solution curve ``bends backwards''. Thus, at a turning point, the forwardly increasing discretization of $\alpha$ would lead to losing a part of the curve. Figure \ref{fig_return_point} shows the result that would be obtained by applying the described continuation method in the presence of a turning point.
\\
To solve this problem, we reparameterize the solution curves by considering the arc length $ (x_n, y_n, \alpha_n)$ for a certain $n$. Then, we include a new circumference equation of adaptive radius $\kappa_n$ in the nonlinear system of algebraic equations \Eqref{ecuacion_catalogue}, which corresponds to intersecting at each iteration the solution curve with the adaptative circumference. 
Intuitively, the proposed solution to deal with the existence of turning points is plotted in Figure \ref{fig_return_point2}, where we denote by $ \chi_i :=  (x_i, y_i, \alpha_i)$ for a certain iteration $i$.
\vspace{-0.2 cm}
\\
Note that as previously done with the adaptive step size, an adaptive radius is considered here in order to speed up the computations. Due to the double multiplicity of solutions of the new systems of equations, we also define the search direction of the solution as an initial condition for the Newton's method:
\begin{equation} \label{cond_inicialNewton}
    (x_0, y_0, \alpha_0) = (x_n, y_n, \alpha_n) + \kappa_n \frac{(x_n - x_{n-1}, y_n - y_{n-1}, \alpha_n - \alpha_{n-1})}{\left[ (x_n - x_{n-1})^2 + (y_n - y_{n-1})^2 + (\alpha_n - \alpha_{n-1})^2 \right]^{1/2}}.
\end{equation} 
\section{Criteria for selection of non-eclipsed points}
\begin{figure}[h!]
\begin{center}
    \includegraphics[scale=0.18]{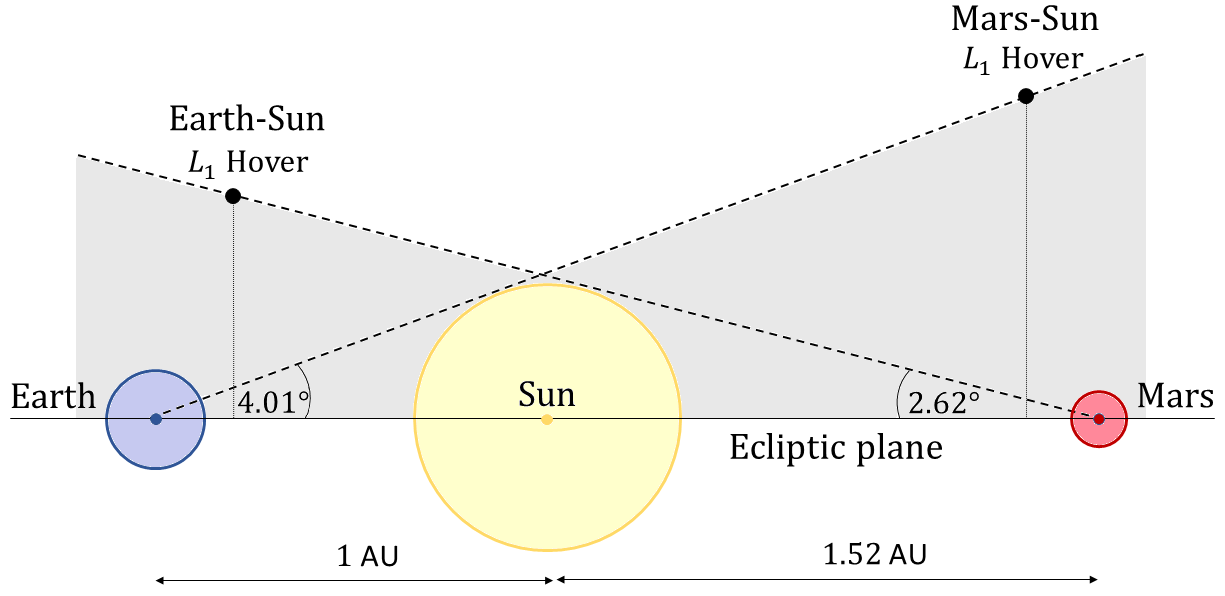}\hspace{0.1 cm}
    \includegraphics[scale=0.27]{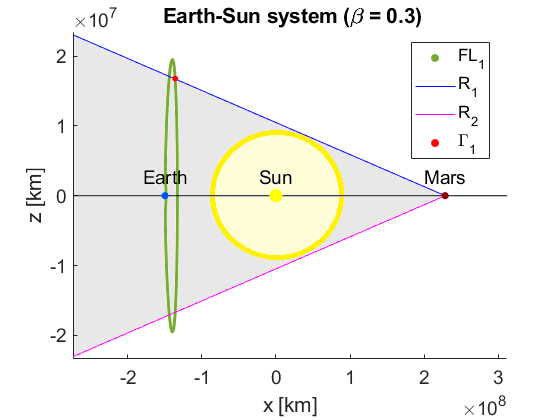}\hspace{0.13 cm}
    \includegraphics[scale=0.27]{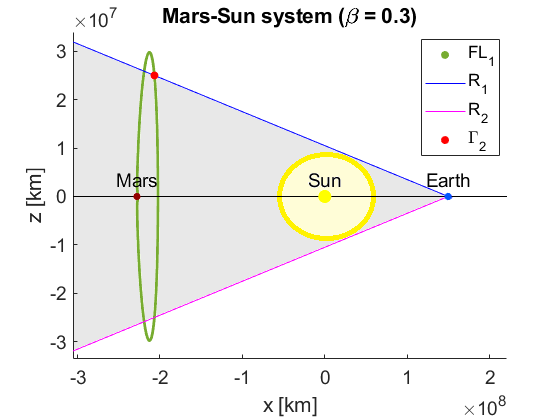}\hspace{0.13 cm}
    \caption{Occultation regions during solar conjunction}
    \label{fig_configuration_eclipsed_regions}
    \end{center}
\end{figure}

During periods of Earth-Sun-Mars alignment, direct communications between the two planets are not feasible. However, for certain values of $\beta$ and $\alpha$, there exist equilibrium points on the $FL_1$ family that do not remain eclipsed. These non-eclipsed equilibrium points lie outside two cones, whose vertices are placed on each of the planets and are tangent to the Sun. On a two-dimensional elevation projection, these cones are delimited by the two half-lines, denoted by $R_1$ and $R_2$, starting on each planet and tangent to the circular projection of the Sun. These half lines enclose the region of the eclipsed points with the minimum elevation angles from each of the planets in the ecliptic plane, (with $z = 0$) and also above or below the ecliptic plane, (with $y = 0$)\cite{Macdonald}. The two admissible non-eclipsed equilibrium points are chosen to lie above these half lines in each of the Earth-Sun and Mars-Sun systems, where we propose to place a CubeSat satellite equipped with a solar sail in order to have a clear view of the other planet. 
\\\\
As a result, in the $xy$ plane, we have to take $\beta \geq 0.05$ for the Earth-Sun system and $\beta \geq 0.02$ for the Mars-Sun, to find equilibrium points of the $FL_1$ family that are no longer eclipsed by the Sun. The linear dynamics of these non-eclipsed equilibrium  points of $FL_1$ is of type outward spiral $\times$ inward spiral, where the growth rate of the outward spiral is significantly small, taking up to years to double its distance to the equilibrium point. 
However, the motion of the planets in the $xy$ plane may cause that these equilibria of the $FL_1$ family become eclipsed by the Sun while in the $xz$ plane, this motion does not affect the equilibria as they only lie above and below the ecliptic plane. Therefore, in the $xz$ plane, we need to consider $\beta \geq 0.3$ to have non-eclipsed equilibrium points on the family $FL_1$. Figure \ref{fig_configuration_eclipsed_regions} shows the configuration of the problem considered for communication between Earth and Mars (left-plot). Moreover, in Figure \ref{fig_configuration_eclipsed_regions} the occultation regions of the family of equilibria $FL_1$, colored in gray, in the $xz$ plane for the Earth-Sun and Mars-Sun systems in dimensional units are also plotted (middle and right plots respectively). Note that larger value of the Sun's radius has been considered due to the presence of solar flares on its surface.
\\\\
Hence, we choose the equilibrium points (red dots in Figure \ref{fig_configuration_eclipsed_regions}) above the half line, $R_1$, corresponding to:
\begin{flalign*}
\Gamma_1 &= ( -1.3589 \cdot 10^{8} , 0, 1.6779 \cdot 10^7 ) \hspace{0.1 cm}\text{km, with} \hspace{0.1 cm} (\alpha, \delta) = (0.3645, 0) \hspace{0.1 cm} \text{rad, for the Earth-Sun system,} \\
\Gamma_2 &= (-2.0675 \cdot 10^{8}, 0,  2.4971 \cdot 10^{7} ) \hspace{0.1 cm} \text{km, with}\hspace{0.1 cm} (\alpha, \delta) = (0.3498, 0) \hspace{0.1 cm} \text{rad, for the Mars-Sun system.} 
\end{flalign*}
Similarly, the symmetric equilibrium points with respect to the $xz$ plane could also have been chosen. The linear dynamics for these points are of the type saddle $\times$ center  $\times$ center where the instability is mainly due to the existence of the unstable manifold associated to the saddle behavior, hence station-keeping maneuvers are required to prevent the satellite from drifting away from the vicinity of the equilibrium points. 
\section{Control strategy for station-keeping around equilibria}
Generally, let the unstable equilibrium point $p_0 = (x_0, 0, z_0, 0,0,0)$ be the fixed point we want to remain close to for a given $\alpha_0$ and $\delta_0$. The control strategy relies on successively rectifying the sail orientation in order to avoid the satellite from escaping a certain threshold distance from the nominal equilibrium point along the direction of the unstable manifold. 
Note that the sail reorientation causes the position of the equilibrium point to be shifted, as well as their stable and unstable manifolds. 
Then, the unstable manifold of the new equilibrium point $p_1 = (x_1, 0, z_1, 0,0,0)$, with the new sail orientation parameters $\alpha_1$ and $\delta_1$, will rectify the trajectory, bringing the satellite back close to the nominal position of the equilibrium point $p_0$. Therefore, the main idea is based on taking advantage of the fact that the phase space of the system changes each time the orientation of the solar sail is changed.\cite{Farres, simo1987optimal, inbook, articleJFarres} 
\\\\
For this purpose, a reference frame $\mathcal{R}$ is considered with the origin at the equilibrium point $p_0$ and the basis of eigenvectors $\mathcal{B} = \lbrace \vec{v_1},\vec{v_2}, \vec{v_3}, \vec{v_4}, \vec{v_5}, \vec{v_6} \rbrace $, where $\vec{v_1}$ and $\vec{v_2}$ are the eigenvectors associated to $\pm \gamma$ respectively; $\vec{v_3}$ and $\vec{v_4}$ the real and imaginary part of the eigenvectors associated to $\pm \eta i$ respectively; and finally $\vec{v_5}$ and $\vec{v_6}$ the real and imaginary part of the eigenvectors associated to $\pm \nu i$ respectively. In this reference frame, $\vec{v_1}$, $\vec{v_2}$ represent the directions of the saddle plane axes, while $\vec{v_3}$, $\vec{v_4}$ and $\vec{v_5}$, $\vec{v_6}$ correspond to the directions of the center plane axes. The projection of the satellite's trajectory on these three planes describes the saddle and the centers behavior of the motion. 
Note that in this reference system, 
it is shown a decomposition of the trajectories into their characteristic shapes where the satellite's motion can be easily described. Moreover, in this new reference system $\mathcal{R} = \lbrace p_0; \mathcal{B}  \rbrace$, we denote by $s_0 = (s_0^1, \dots, s_0^6) = (0, \dots, 0)$ and $s_1 = (s_1^1, \dots, s_1^6)$ the coordinates of the equilibrium points $p_0$ and $p_1$, where $s_0^p = (s_0^1, s_0^2) = (0,0)$ and $s_1^p = (s_1^1, s_1^2)$ correspond only to the projections of $s_0$ and $s_1$ on the saddle plane respectively. 
\\\\
Locally, in the saddle plane, for $(\alpha_0,\delta_0)$ the trajectories can be approximated by:
\begin{flalign}
s^1_{out}(t) &\approx \Bar{s}^1_0 e^{\gamma (t -t_0)}, \label{cambiovela_1} \\
s^2_{out}(t) &\approx \Bar{s}^2_0  e^{-\gamma (t -t_0)},  \nonumber
\end{flalign}
where $(\Bar{s}^1_0, \Bar{s}^2_0)$ is the initial point of the trajectory at $t = t_0$. From now on we refer to these trajectories as the outward motion as it is related to the trajectories escaping from the vicinity of $p_0$. In the reference system $\mathcal{R}$, we denote the trajectory of the outward motion by $(s^1_{out}(t), s^2_{out}(t), s^3_{out}(t), s^4_{out}(t), s^5_{out}(t), s^6_{out}(t))$, where $(s^1_{out}(t), s^2_{out}(t))$ describes the saddle behavior while the pairs $(s^3_{out}(t), s^4_{out}(t))$ and $(s^5_{out}(t), s^6_{out}(t))$ the central behavior.
\\\\
When at a certain $t_1>t_0$, the sail orientation changes, new solar sail parameters $\alpha_1$ and $\delta_1$ are considered for the equilibrium point $p_1$. Then, the trajectories in the saddle plane are approximated as: 
\begin{flalign}
s^1_{in}(t) &\approx s_1^1 + (\Tilde{s}^1_0 - s_1^1)e^{\Bar{\gamma} (t -t_1)}, \label{cambiovela_2} \\
s^2_{in}(t) &\approx s_1^2 + (\Tilde{s}^2_0 - s_1^2)e^{-\Bar{\gamma} (t -t_1)}, \nonumber
\end{flalign}
where $(\Tilde{s}^1_0, \Tilde{s}^2_0)$ are the initial conditions and $\pm \bar{\gamma}$ the real eigenvalues for $p_1$. Similarly, the trajectories approaching the equilibrium point $p_0$ correspond to the inward motion. Once again, the trajectory of the inward motion expressed in the reference system $\mathcal{R}$ is $(s^1_{in}(t), s^2_{in}(t), s^3_{in}(t), s^4_{in}(t), s^5_{in}(t), s^6_{in}(t))$. After that, we restore the initial orientation angles (i.e. $\alpha_0$ and $\delta_0$) of the nominal equilibrium point $p_0$. Then, the trajectory escapes again, so the process described above is repeated until the final mission time. 
\\\\
The determination of the new equilibrium point is crucial in this strategy. As mentioned before, the appropriate calculation of its solar sail orientation parameters will rectify the satellite's trajectory. 
\subsection{Conditions for the control strategy}
\begin{figure}[h!]
    \centering
    \includegraphics[scale = 0.29]{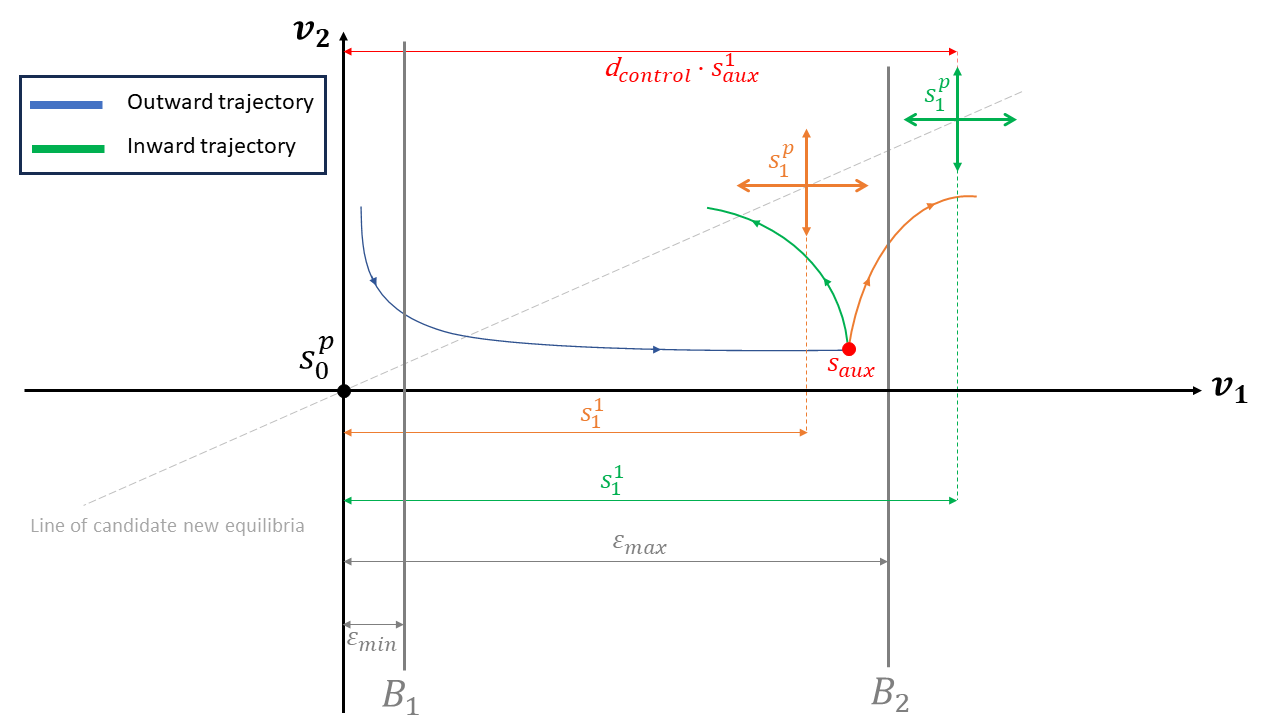}
    \caption{Schematic idea of the control strategy in the saddle plane}
    \label{fig_trayectorias_posibles}
\end{figure}

The unstable manifold in the saddle plane is the one that cause the satellite's trajectory to escape. Then, to ensure that the satellite's trajectory always remains close to $p_0$, we define a region of movement in the saddle plane by using two boundaries, $B_1$ and $B_2$, as:
\begin{flalign*}
B_1 &= \lbrace (s^1,s^2): | s^1 - s_0^1 | = \varepsilon_{\min} \rbrace, \\
B_2 &= \lbrace (s^1,s^2): | s^1 - s_0^2 | = \varepsilon_{\max} \rbrace,
\end{flalign*}
where $\varepsilon_{\max} > \varepsilon_{\min} > 0$ are the maximal and minimal distances of the trajectories to the projection from the nominal equilibrium point, $p_0$. Then, when the trajectory reaches one of these boundaries we must maneuver (by changing the sail orientation) to ensure that the motion remains controlled. In particular, if the initial orientation of the sail is set for $ \alpha = \alpha_0$ and $\delta = \delta_0$, the trajectory follows the dynamics of \Eqref{cambiovela_1}, going from $B_1$ to $B_2$ as it escapes from $p_0$. When we maneuver, the trajectory obeys equations \Eqref{cambiovela_2}, so that if the values of $\alpha_1$ and $\delta_1$ are chosen appropriately, the motion will be directed from $B_2$ to $B_1$.
\\\\
Hence, a key condition is imposed for the control strategy. The new equilibrium point $s_1^p = (s_1^1, s_1^2)$ in the saddle plane must fulfill $\abs{s^1_1} > \varepsilon_{\max}$, to ensure that the trajectory will approach the stable manifold of the equilibrium point $p_0$. With this condition, we have a behavior like the one colored in green in Figure \ref{fig_trayectorias_posibles}. Otherwise, if $s_1^p$ is on the left-hand side of $B_2$, the trajectory does not return to $s_0^p$, instead it escapes in the direction of the saddle of the new equilibrium point $s_1^p$ as depicted in orange in Figure \ref{fig_trayectorias_posibles}. Note that the dotted straigh line (colored in gray) represents the possible distribution of the new equilibrium points along $s_1^p$ for different values of $\alpha$ and $\delta$. 
\\
Therefore, to obtain the new equilibrium point $s_1^p$ outside the $B_2$ boundary, we define an auxiliary point $s_{aux}$ on the outward trajectory that is sufficiently close to the $B_2$ boundary, verifying that its first coordinate is of the form $ |s_{aux}^1 |= \varepsilon_{max} - \varepsilon$, for a certain $0 \leq \varepsilon \ll 1$. Since the equations of motion are solved numerically over a finite set of nodes, $s_{aux}$ can be taken as the last point obtained before surpassing the boundary $B_2$. This $s_{aux}$ is the maneuver point, where the sail orientation is changed. For this purpose, a new control parameter $d_{control} > 1$ is defined, which corresponds to the scaling factor by which $ s_{aux}^1$ should be multiplied to ensure that $ \abs{s_{1}^1} > \varepsilon_{max}$. Consequently, the desired equilibrium point position $s_1$ is \cite{Farres}:
$$ s_1 = \left(d_{control} \cdot s_{aux}^1,  \hspace{0.1 cm}  \frac{1}{2}s_{aux}^2,  \hspace{0.1 cm} \frac{1}{2} s_{aux}^3,  \hspace{0.1 cm} \frac{1}{2}s_{aux}^4, \hspace{0.1 cm}  \frac{1}{2}s_{aux}^5,  \hspace{0.1 cm} \frac{1}{2}s_{aux}^6 \right)^T,  \hspace{0.2 cm} \text{verifying} \hspace{0.2 cm}  \abs{d_{control} \cdot s_{aux}^1} > \varepsilon_{\max}.$$
We have considered the $1/2$ factor to ensure that the center behavior decreases, or at least remains bounded. The appropriate value of $d_{control}$ should not be excessively large, since the linearized dynamics would no longer represent the system reliably. Nevertheless, within a range of admissible values, a larger value of $d_{control}$ allows the trajectory to return more rapidly to $s_0^p$. 
\\\\
Furthermore, an estimate of the time that the satellite takes to go from one boundary to another can be obtained. For instance, from $B_1$ to $B_2$, 
denoting by $\Delta t_1$ the time it takes to get from $\Bar{s}^1_0  = \varepsilon_{\min}$ to $s^1_{out}(T) = \varepsilon_{\max}$ and following the linear dynamics \Eqref{cambiovela_1}, we obtain:
\begin{equation} \label{DeB1aB2}
    \Delta t_1 = \frac{1}{ \lambda} \log \left( \frac{\varepsilon_{\max}}{\varepsilon_{\min}}  \right).
\end{equation}
Likewise, from $B_2$ to $B_1$, the inward trajectory follows \Eqref{cambiovela_2}, going from $\Tilde{s}^1_0 = \varepsilon_{\max}$ to $s^1_{in} = \varepsilon_{\min}$. Then, we have:
\begin{equation} \label{DeB2aB1}
    \Delta t_2 = \frac{1}{ \bar{\lambda}} \log \left( \frac{ s_1^1 -\varepsilon_{\min}}{ s^1_1 - \varepsilon_{\max}}  \right).
\end{equation}
Note that $\Delta t_2$ varies depending on the election of the position of the new equilibrium point. Particularly, the further away the new equilibrium point position is, the less time it takes to return close to $s_0^p$. 
\subsection{Determination of the new sail orientation}
Once the location of the desired equilibrium point $p_1$ is determined, we seek for the associated values of $\alpha_1$ and $\delta_1$. Since $p_0$ is an equilibrium point, it satisfies $F(p_0, \alpha_0, \delta_0) = 0$, where $F$ is as previously, obtained equating the right-hand side of \Eqref{ecuacion_catalogue} to zero. Recall that we do not have an explicit expression for the equilibrium points $p(\alpha, \delta)$. However, by means of the Implicit Function Theorem,
if  $DF(p_0, \delta_0, \alpha_0)$ is non-singular, there is a local mapping $p = p(\alpha, \delta)$ (whose expression is still unknown) such that $F(p(\alpha, \delta), \alpha, \delta) = 0$. Then, the derivatives of $p$ must satisfy:
\begin{flalign}
DF(p_0, \delta_0, \alpha_0) \frac{\partial p}{\partial \alpha} (\alpha_0, \delta_0) &= - \frac{\partial F}{\partial \alpha} (p_0, \alpha_0, \delta_0), \label{derivada_p1}\\
DF(p_0, \delta_0, \alpha_0) \frac{\partial p}{\partial \delta} (\alpha_0, \delta_0) &= - \frac{\partial F}{\partial \delta} (p_0, \alpha_0, \delta_0), \nonumber
\end{flalign}
where $DF$ is the Jacobian matrix of $F$. 
\\\\
As the new fixed point is close enough to the nominal equilibrium point $p_0$ in order to avoid that the trajectory escapes, $\alpha_1$ and $\delta_1$ will be also close to the values of $\alpha_0$ and $\delta_0$. Consequently, we can consider the general linear approximation of $p$ by neglecting terms of quadratic or higher order in the Taylor expansion:
\begin{equation} \label{expresion_p_sin_explicitar}
    p(\alpha, \delta) = p(\alpha_0, \delta_0) + Dp \cdot h, \hspace{0.2 cm} \text{with} \hspace{0.2 cm} Dp = \left( \frac{\partial p}{\partial \alpha } (\alpha_0, \delta_0), \frac{\partial p}{\partial \delta } (\alpha_0, \delta_0) \right),
\end{equation}
where we seek for $ h$ defined as:
\begin{equation} \label{expresion_h_Dp} 
    h := (h_1, h_2) = (\alpha - \alpha_0, \delta - \delta_0)^T. 
\end{equation}
Evaluating the expression \Eqref{expresion_p_sin_explicitar} into the case $\alpha = \alpha_1$ and $\delta = \delta_1$, we have:
\begin{equation} \label{expresion_p_sin_explicitar2}
    p_1 = p_0 + Dp \begin{pmatrix}
    \alpha_1- \alpha_0 \\
    \delta_1 - \delta_0
    \end{pmatrix}.
\end{equation}
Let $M_v$ be the change basis matrix formed by the vectors of $\mathcal{B}$, then \Eqref{expresion_p_sin_explicitar2} is rewritten in the reference system $\mathcal{R}$ as:
\begin{equation} \label{sistema_6ecuaciones_2incognitas}
    s_1 =  M_v^{-1} Dp \cdot h =: A \cdot h.
\end{equation}
For a certain $s_1$ that we specify, if we find the values of $\frac{\partial p}{\partial \alpha } (\alpha_0, \delta_0)$ and $\frac{\partial p}{\partial \delta } (\alpha_0, \delta_0)$ from solving the linear system of equations \Eqref{derivada_p1}, we have an overdetermined system \Eqref{sistema_6ecuaciones_2incognitas} of 6 equations and 2 unknowns, $\alpha_1$ and $\delta_1$. This system does not admit an unique solution, so we obtain its least squares solution. Hence, denoting by $A = (a_{i,j})_{j=1, 2}^{i=1, \dots , 6}$ to the elements of the matrix $A$, we distinguish the following cases:
\begin{itemize}
    \item If $a_{11} = a_{12} = 0$: $\frac{\partial p}{\partial \alpha}$ and $\frac{\partial p}{\partial \delta}$ are orthogonal to $\vec{v_1}$. In this case, the trajectory cannot be controlled using this strategy, since there are no fixed points which bring the trajectory back close to the nominal equilibrium point $p_0$. 
    \item If both are not null, we can control the system. Then, in order to reduce the system, a possible option can be obtained one of the unknowns $h_1$ or $h_2$ as a function of the other. Thus, fixing $s_1^1$, we have $s_1^1 = a_{11}h_1 + a_{12}h_2$. Since at least one of the elements $a_{11}$ or $a_{12}$ is nonzero, we can isolate $h_1$ or $h_2$. For numerical stability when dividing, we take the one with the maximum absolute value. Furthermore, to also ensure $| s_1^1 | > \varepsilon_{\max}$, we consider:
    \begin{enumerate}
    \item If $|a_{11}| = \max(|a_{11}|, |a_{12}|)$: 
    \begin{equation} \label{caso1}
        s_1^1 = a_{11}h_1 + a_{12}h_2 \Rightarrow h_1 = \frac{s_1^1 - a_{12}h_2}{a_{11}}.
    \end{equation}
    \item If $|a_{12}| = \max(|a_{11}|, |a_{12}|)$: 
    \begin{equation} \label{caso2}
        s_1^1 = a_{11}h_1 + a_{12}h_2 \Rightarrow h_2 = \frac{s_1^1 - a_{11}h_1}{a_{12}}.
    \end{equation}
    \end{enumerate}
\end{itemize}
Thus, in either case 1 or 2, the system \Eqref{sistema_6ecuaciones_2incognitas} is now reduced to a system of 5 equations and 1 unknown, which we denote in general by:
\begin{equation} \label{sistema_final_min2}
    \bar{s_1} = \bar{A} \cdot \bar{h}.
\end{equation}
Consequently, solving the system  \Eqref{sistema_final_min2} by means of least squares approximation, we obtain the $\widehat{\bar{h}}$ such that $|| \bar{s_1} - \bar{A} \cdot \widehat{ \bar{h}} || $ is minimum. Therefore, the least squares estimation of $\bar{h}$ is given by:
\begin{equation}
    \widehat{ \bar{h}} = (\bar{A}^T \bar{A})^{-1} \bar{A}^T \bar{s_1}.
\end{equation}
Hence, depending on the case, once we have obtained $h_1$ or $h_2$, (i.e. $\alpha_1$ or $\delta_1$), the other sail orientation parameter is computed using \Eqref{caso1} or \Eqref{caso2}. Note that the linear dynamics is only used to find the suitable new sail orientation. For the mission simulations, we use the whole system of equations.
\section{Mission simulations}
In this section, numerical results are shown for a mission lifetime $T = 600$ days where the control strategy described in the previous section is applied. The control parameters considered for both the Earth-Sun and Mars-Sun systems are given in Table \ref{tab:311_control}. Note that these parameters depend essentially on the dynamical properties as well as the mission purposes. However, they have been chosen to provide an accurate description of the motion.   

\begin{figure}[h!]
    \centering
       \textbf{Earth-Sun system}\par\medskip
    \includegraphics[scale = 0.33]{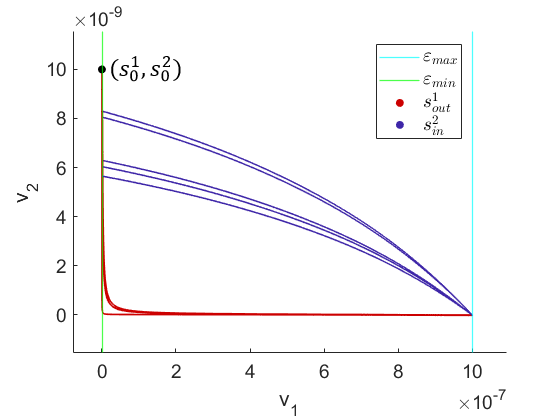} \hspace{0.1cm}
     \includegraphics[scale = 0.33]{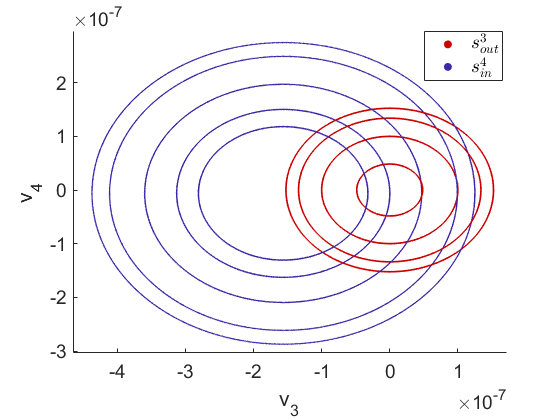} \hspace{0.1 cm}
      \includegraphics[scale = 0.33]{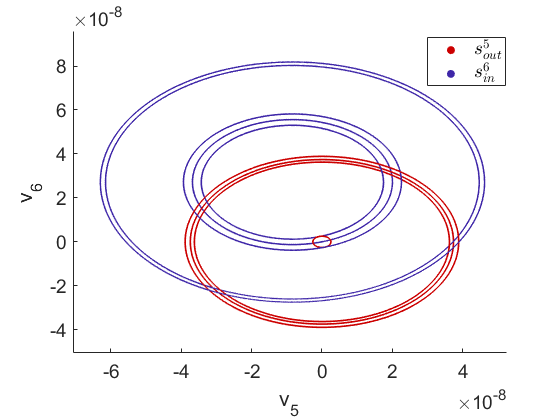}
           \\[\smallskipamount]
         \textbf{Mars-Sun system}\par\medskip 
         \includegraphics[scale = 0.33]{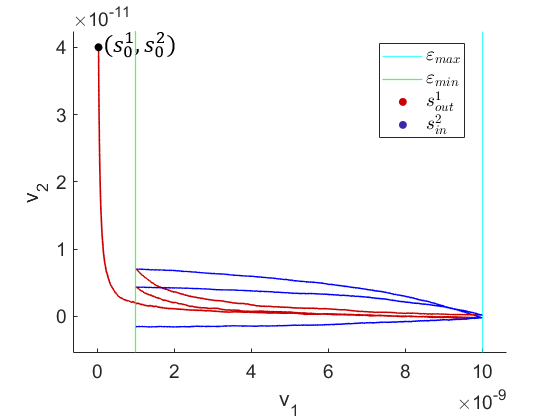} \hspace{0.01 cm}
     \includegraphics[scale = 0.33]{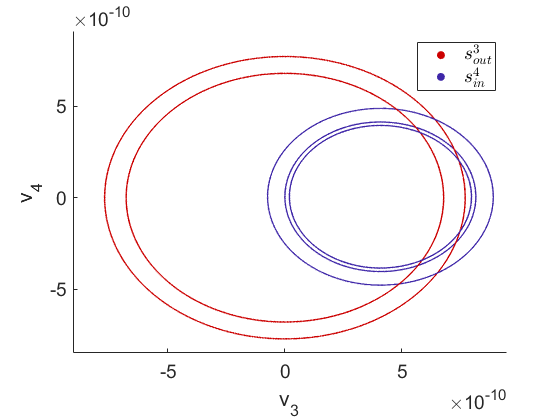} \hspace{0.01 cm}
      \includegraphics[scale = 0.33]{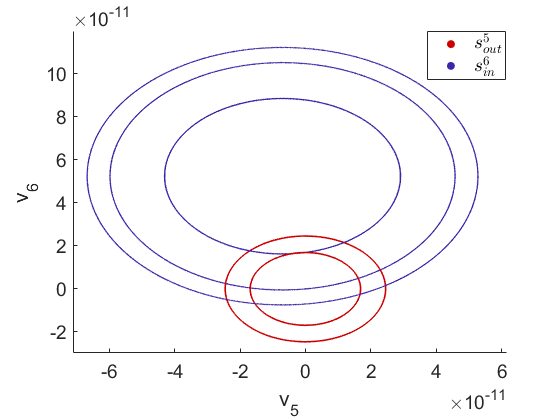}
    \caption{Control strategy in the saddle and the two center projections}
    \label{FIG_CONTROL_ptos_sillacentrocentro}
\end{figure}

\begin{figure}[h!]
    \centering
           \textbf{Earth-Sun system}\par\medskip
    \includegraphics[scale = 0.35]{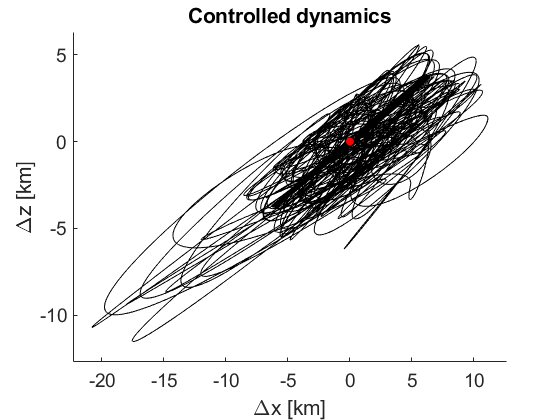} \hspace{1 cm}
     \includegraphics[scale = 0.24]{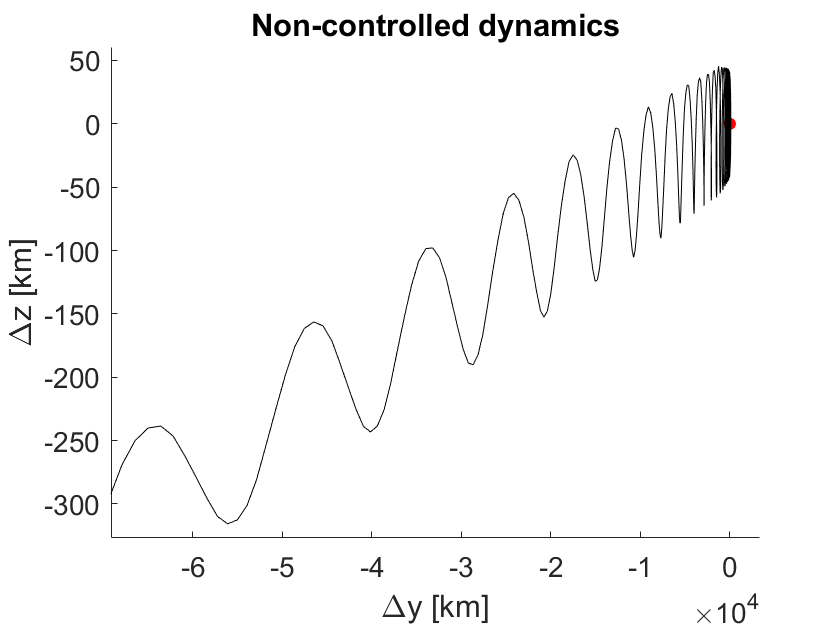} 
    \caption{Dynamics in the Earth-Sun system centered in the non-eclipsed equilibrium point}
        \label{fig_control_nocontrol_tierra}
\end{figure}

\begin{figure}[h!]
    \centering
           \textbf{Mars-Sun system}\par\medskip
    \includegraphics[scale = 0.35]{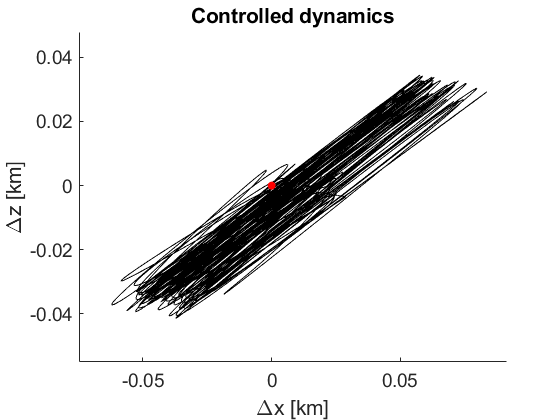} \hspace{1 cm}
     \includegraphics[scale = 0.24]{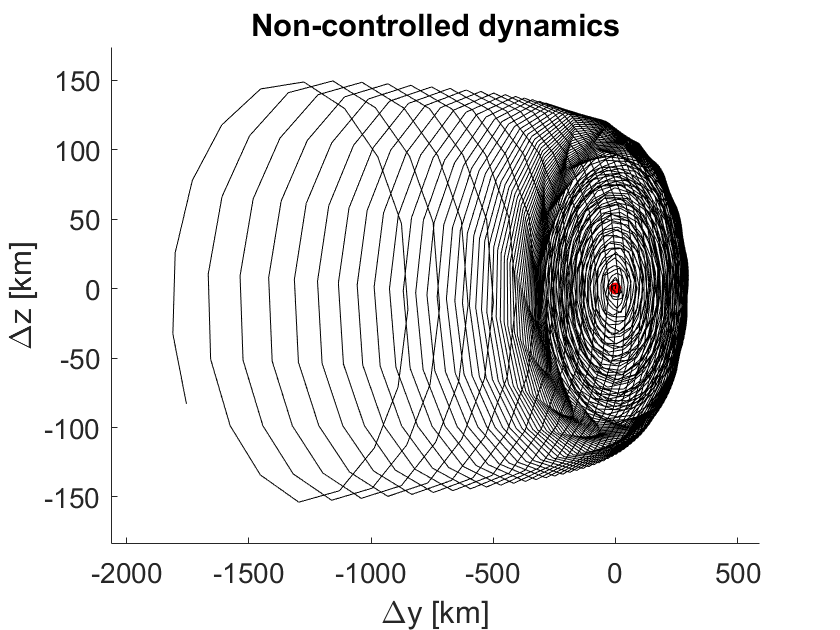} 
    \caption{Dynamics in the Mars-Sun system centered in the non-eclipsed equilibrium point}
        \label{fig_control_nocontrol_Marte}
\end{figure}

\begin{figure}[h!]
    \centering
    \includegraphics[scale = 0.36]{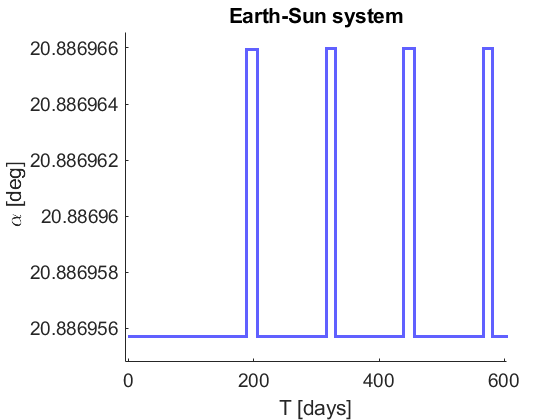} \hspace{1 cm}
     \includegraphics[scale = 0.36]{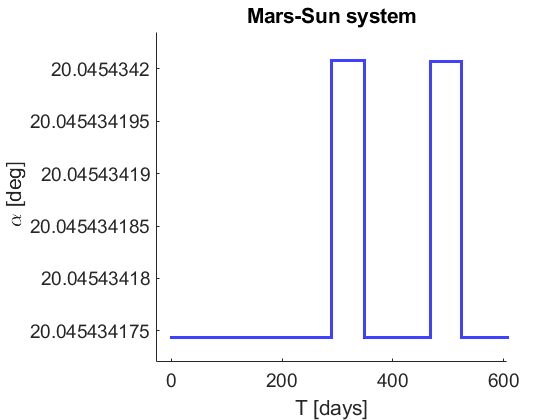} 
    \caption{Values of $\alpha$ obtained by applying the control strategy for $T = 600$ days}
        \label{fig_alpha_variations}
\end{figure}

\newcolumntype{g}{>{\columncolor{LightCyan}}c}
\begin{table}[ht]
\centering
\begin{tabular}{g| c| c| c} 
\hline
\rowcolor{Gray}
$ $ &  $\varepsilon_{\max}$& $\varepsilon_{\min}$ & $d_{control}$ \\ 
\hline
Earth &$ 10^{-6}$ & $10^{-9}$ & 2 \\
Mars &  $10^{-8}$ & $10^{-9}$ & 1.5 \\ 
\hline
\end{tabular}
\caption{Control parameters for Earth and Mars}
\label{tab:311_control}
\end{table}
\newpage 
\noindent Figure \ref{FIG_CONTROL_ptos_sillacentrocentro} shows the projection of the trajectory on the saddle and two center planes for both the Earth-Sun and the Mars-Sun systems.
\\
As a result, the outward and inward sail trajectories described in the saddle plane remain closed to the nominal equilibrium point due to the boundaries $B_1$ and $B_2$ while a sequence of rotations described by this sail around the different new equilibrium points are performed on the center planes. The composition of these rotations is also bounded without a significant growth of the central motion. Notice that the outward trajectory is colored in red while the inward one is colored in blue. The differences between the saddle behaviors of both systems are mainly due to the distribution of the new admissible equilibrium points in each iteration. We must mentioned that, establishing larger distances between the boundaries for the Mars-Sun case may cause the system to not follow its linearized dynamics.
\\\\
In Figures \ref{fig_control_nocontrol_tierra} and \ref{fig_control_nocontrol_Marte}, we find represented the controlled (left) and uncontrolled (right) trajectories centered in the equilibrium points (colored in red) for both the Earth-Sun and the Mars-Sun systems in rescaled dimensional units. Note that the uncontrolled trajectory is depicted in the $yz$ projection for a better comprehension. Without a control strategy, a satellite ---governed by the gravitational dynamics of the RTBP including the SRP effect--- would quickly destabilize, moving away from the equilibrium point and thus being totally inoperative for the purpose of establishing communications between both planets. Note that the satellite experiences larger displacements from the equilibrium point in the Earth-Sun system due to the difference imposed in the orders between $\varepsilon_{\min}$ and $\varepsilon_{\max}$. Finally, the variations in the angle orientation $\alpha$ for both Earth-Sun and Mars-Sun systems are plotted in Figure \ref{fig_alpha_variations}. 
\\
\\
\\
\noindent We notice that the trajectory escapes faster in the Earth-Sun system than in Mars-Sun system. This is a consequence of the positive saddle eigenvalue being greater for the Earth-Sun case, hence having faster escape rates. As a result, for the Mars-Sun system, this slow motion causes less required changes in the sail orientation to maintain the satellite close to $p_0$ during the lifetime of the mission. Table \ref{tab:eigenvaluess} shows the values of the eigenvalues for both equilibrium points $\Gamma_1$ and $\Gamma_2$.
\begin{table}[h!]
\begin{center}
\begin{tabular}{|c| c| c| c| c | c | c} 
\hline
 Equilibria & $ \pm \gamma$ & $\pm \eta$ & $\pm \nu$  \\ 
\hline
$\Gamma_1$ &$  \pm 0.056925035468768 $ & $ \pm 1.086660790998318i$ & $\pm 0.915119124576269i$ \\
$\Gamma_2$  &  $ \pm 0.018920715025163$ &  $ \pm 1.084847855035612i$ &  $ \pm 0.915337229119532i$ \\ 
\hline
\end{tabular}
\caption{Eigenvalues for the equilibrium points $\Gamma_1$ and $\Gamma_2$}
\label{tab:eigenvaluess}
\end{center}
\end{table}\\\\
Notice that, the variations of $\alpha$ in both cases are very small. This behavior is mainly due to the limitations in the selection of the control parameters and the strong dependence between them. Firstly, in each iteration, due to the distributions of the new equilibrium points $s_1^p$ in the saddle projection, to obtain larger variations of the sail orientation angles would require taking the new equilibria at greater distances $d_{control}$. However, for greater distances, the linearized dynamics no longer approximate the trajectories accurately. For this same reason, the boundaries $B_1$ and $B_2$ have been selected close enough to the nominal equilibrium point $s_0^p$. 
\\\\
In fact, depending on the distance at which we place one boundary from the other, we allow the trajectory to propagate further, thus increasing the time between maneuvers at which the orientation angles are kept fixed. Therefore, in both cases, the new sail orientation angles remain fixed for months of mission period while the initial orientation angles $\alpha_0$ and $\delta_0$ last less. The reason is that the inward trajectory goes faster than the outward one due to the set value of $d_{control}$. Finally, it should be noted that the results are independent of the initial condition when taken as a slight perturbation of the position of the equilibrium point $s_0$. The selection of these initial conditions only affects the initial location of the trajectory in the saddle branches.
\\\\
Overall, in both cases, the bounds imposed on the trajectories defined by the control parameters $\varepsilon_{\max}$, $\varepsilon_{\min}$ and $d_{control}$ on the saddle planes reflects the control of the satellites' trajectories near the equilibrium points $\Gamma_1$ and $\Gamma_2$ for the respective Earth-Sun and Mars-Sun systems.
\section{Conclusions}
In this paper, the analysis of a mission to establish communications between Earth-Mars during solar conjunctions has been performed. 
First, we numerically determined the equilibrium points of the system through an iterative continuation method. This numerical method relied on fixing $\delta$ and varying the value of the parameter $\alpha$ in $[-\pi/2, \pi/2]$. Moreover, the method was adapted to solve the problem of the existence of turning points. Then, we considered the invariant $xy$ plane (with $\delta = \pi/2$) and the $xz$ plane (with $\delta=0$). As a result, from all the feasible families of equilibrium points obtained for both Earth-Sun and Mars-Sun systems, we determined the suitable equilibrium points of the family $FL_1$ that were not eclipsed in both systems.
\\\\
These equilibrium points are unstable, hence it was necessary to use station-keeping maneuvers in order to control the saddle behavior. 
The aim of the control strategy that we have considered in this paper consists of maintaining the sail's trajectory close to the unstable equilibrium point by taking advantage of the properties of the systems natural dynamics. To do so, we appropriately change the sail orientation depending on some limitations so that the satellite returns close to the point, then avoiding that the satellite's trajectory escapes in the direction of the unstable manifold.
\\\\
With the implementation of this control strategy algorithm, numerical simulations were performed in both Earth-Sun and Mars-Sun systems to show its effectiveness. 
Finally, the methods presented in this paper successfully address the complex problem of controlling the orbital dynamics of a satellite equipped with a solar sail near certain unstable non-eclipsed equilibrium points for the case of an Earth-Sun-Mars alignment, thus proposing a solution to the existing communication problem between them. 
\\\\
For future work, given that the changes in sail orientation required to control the satellite are small, we will study other control options that allow larger changes on the sail orientation, for more realistic scenarios. This approach may be accomplish by focusing on the non-linear dynamics of the problem.

\section{ACKNOWLEDGEMENTs}
This work has been supported by the Spanish grant BOE-B-2022-21646, awarded and funded by the National Geographic Institute, under the Public Ministry of Spain. 
\bibliographystyle{AAS_publication}   
\bibliography{AAStemplate}   
\end{document}